\documentclass [aps,prL,twocolumn,showpacs,longbibliography]{revtex4-1}

\usepackage{amsmath}
\usepackage{amssymb}
\usepackage{amsthm}
\usepackage{mathrsfs,mathbbol}
\usepackage{graphicx}
\usepackage{verbatim}
\usepackage{bm}
\usepackage{color}
\usepackage[dvipsnames]{xcolor}
\usepackage{graphicx,epsfig,amsfonts,amssymb}

\usepackage[colorlinks,
linkcolor=blue,
anchorcolor=blue,
urlcolor=blue,
citecolor=blue
]{hyperref}

\usepackage{bm}
\usepackage{times}

\usepackage[all]{xy}

\newcommand{\sgn}{\operatorname{sgn}}

\newcommand{\be}{\begin{eqnarray}}
\newcommand{\ee}{\end{eqnarray}}

\begin{document}

\title{Loop unitary and phase band topological invariant in generic multi-band Chern insulators}

\author{Xi Wu} 
\email{wuxi@htu.edu.cn}
\affiliation{School of Physics, Henan Normal University, Xinxiang 453007, China}
\author{Ze Yang} 
\author{Fuxiang Li}
\affiliation{School of Physics and Electronics, Hunan University, Changsha 410082, China}
\begin{abstract}
	Quench dynamics of topological phases has been studied in the past few years and dynamical topological invariants are formulated in different ways. Yet most of these invariants are limited to minimal systems in which Hamiltonians are expanded by Gamma matrices. Here we generalize the dynamical 3-winding number in two-band systems into  the one in generic multi-band Chern insulators and prove that its value is equal to the difference of Chern numbers between post-quench and pre-quench Hamiltonians. Moreover we obtain an expression of this dynamical 3-winding number represented by gapless fermions in phase bands depending only on the phase and its projectors, so it is generic for the quench of all multi-band Chern insulators. Besides, we obtain a multifold fermion in the phase band in $(\mathbf{k},t)$ space by quenching a three-band model, which  cannot happen for two band models. 
\end{abstract}
\date{\today}
\maketitle
\section{Introduction}
Topological phases, as new phases of matter, were originally proposed in static systems \cite{hasan2010colloquium,Wen:2004ym}. In the past decade, more and more research have been investigated in dynamical systems such as Floquet systems\cite{PhysRevB.82.235114,Lindner:2011aa,PhysRevLett.106.220402,PhysRevX.3.031005,PhysRevLett.110.200403,DAlessio:2015aa,PhysRevX.6.041001,PhysRevLett.123.066403,PhysRevLett.124.057001,PhysRevB.96.155118,PhysRevB.96.195303,PhysRevB.100.085308,Li:2023aa,Jangjan:2020aa,PhysRevB.106.224306} under a periodic driving Hamiltonian in which topological invariants are studied. Besides, explorations of topological phases after a quench are widely performed in the past years both in experiments\cite{doi:10.1126/science.aad4568,PhysRevLett.107.235301,PhysRevLett.113.045303} and in theory\cite{CaioMD2015,PhysRevB.94.155104,Ying2016,Tarnowski:2019aa,PhysRevLett.125.053601}. {After preparing a ground state of a quantum system, one suddenly changes the Hamiltonian and the ground state wave function will undergo a unitary time-evolution according to this new Hamiltonian. Based on the method of momentum resolved tomography, mapping out the expectation value of an observer such as spin, and tracking its trajectory, one can obtain the information of time-evolution(for two band model in two dimensions, a momentum point leaves a curve on a Bloch sphere). The topology of the trajectory is related with the topology of the pre-quench and post-quench Hamiltonians.}  

Although Chern number after quench\cite{DAlessio:2015aa,PhysRevB.97.195127} was shown to be invariant under time evolution, different approaches have been taken to look for dynamical topological invariants. One approach is the so-called dynamical bulk-edge correspondence\cite{ZHANG20181385,ZhangLong2019,YaWangXJLiu2019,YuXiangLong2021,ZhangLong2020,ShuaiChen2018,LI20211502,PhysRevA.107.052209,PhysRevResearch.2.023043} which constructs topological invariant by the time-averages of observables in the long-time limit, around the so-called band-inversion surfaces.  This approach is successful not only in a sudden quench but also in a slow quench \cite{PhysRevA.102.042209,PhysRevA.106.022219}. There is another approach of constructing dynamical topological invariants which is based on a periodic time-momentum manifold by a rescaling of time. This approach also relates the dynamical topological invariants with static topological invariants. These includes the one for the one dimensional models \cite{PhysRevB.97.060304}  and Hopf invariant \cite{WangCe2017,ChenXin2020,Flaschner:2018aa}, loop unitary and its homotopy invariant\cite{HuHaiping2020,PhysRevB.101.155131} for the quench process of Chern insulators/Hopf insulators in two/three dimensions. 

Dynamical topological invariants proposed in various of dimensions above are mainly for minimal models of topological phases, in which Hamiltonians are expanded by Gamma matrices. The minimal models are the simplest models for Chern insulators. However, in more generic situations in condensed matter physics, multi-band Chern insulators beyond these are encountered\cite{PhysRevLett.109.235308}. 
For a better understanding of the mathematics and the physics, the generalized Bloch spheres were introduced\cite{JAKOBCZYK2001383,Zyczkowski2003,KIMURA2003339,PhysRevA.68.062322,PhysRevB.104.085114}, which showed the complexity of the study of generic multi-band Chern insulators even in the static situation. Due to this complexity, an explicit picture for the quench dynamics of multi-band insulators and formulation of its dynamical topological invariants have been delayed comparing with the minimal models. 

Recently, the time-evolving wave functions of these Chern insulators from quench are successfully mapped out, using the description of nested-sphere \cite{PhysRevResearch.4.023120}. So the quench dynamics for multi-band Chern insulators based on the coherence vector on the generalized Bloch sphere becomes manageable.
It is therefore fully practical and important to construct a dynamical topological invariant for multi-band Chern insulators. Among the dynamical topological invariants mentioned above, the homotopy invariant of the loop unitary proposed in \cite{HuHaiping2020} is a promising candidate to describe the quench process of multi-band insulators because it is defined by the Wess-Zumino-Witten term which is not limited by the form of the Hamiltonian. In the two-band model this dynamical homotopy invariant, measuring the difference of post- and prequench Chern numbers, also manifests as defects in phase bands, which are Weyl fermions in the time-momentum manifold. This is a new picture that relates gapped and gapless fermions in different dimensions which is not originated from the traditional bulk-edge correspondence\cite{PhysRevLett.71.3697,PhysRevResearch.2.043136,PhysRevB.95.165443}. Note that this 3-winding number can be related to 4 dimensional Pontryagin class by a dimensional extension\cite{PhysRevB.109.165125}. {Two natural question arise if we generalize the loop unitary to multi-band insulators:  Will the Wess-Zumino-Witten term give an integer valued homotopy invariant? Will this dynamical homotopy invariant also manifest itself as gapless fermions generically, and will there be multifold fermions as in the static case\cite{PhysRevB.85.155118,PhysRevB.93.045113,BradlynScience2016,PhysRevLett.119.206402,PhysRevLett.119.206401,Schroter:2019aa,Robredo:2024aa}?}

In this work, we answer these two questions. We generalize the loop unitary of Chern insulators into multi-band models by a properly defined time-rescaling for each band. Based on this condition, we prove that the corresponding homotopy invariant is an integer and that its value is equal to the difference of post-quench and pre-quench Chern numbers.  {The strategy for two band model developed in \cite{HuHaiping2020} relays on a relation between Abelian Chern-Simons action and static Chern number in \cite{WangCe2017}, which becomes complicated for multi-band case since the Chern-Simons action becomes non-Abelian. We instead find a more direct approach. This is done in Sec. (\ref{Loopuni}). Moreover when expressing this homotopy invariant as defects of phase bands of the loop unitary, strikingly we find that the contributions from different bands are decoupled from each other and each phase band gives an integer number. Based on this, we find a generic correspondence between static gapped fermions and dynamical gapless fermions after quench.} 
Using eigenprojectors instead of expansion by Pauli matrices, our results can describe quench process of models having any number of bands.  We also provide formulae for Hamiltonians expanded by su(N) algebras. This is done in Sec.(\ref{phasebandchern}). Finally, in Sec.(\ref{Mutilfoldfermion})we find an example that a three-fold fermion appears as the defect of phase band in a quench of a three band model, showing its phase band dispersion and Berry curvature vector field. In Sec.(\ref{SumDisc}) we discuss several topics as future directions.

\section{Loop unitary for quench dynamics in generic multi-band Chern insulators}\label{Loopuni}
In this section, we briefly review the setup and results of \cite{HuHaiping2020} and then show that by an appropriate band-flattening these can work for a generic multi-band insulators. In the end of this section, we also explain that an inappropriate band-flattening may even lead to non-integer-valued 3-winding number.  In \cite{HuHaiping2020} the loop unitary operator
\be\label{LU}
	U_l(t)=e^{-i\mathbf{h}t}e^{i\mathbf{h}_0t}
\ee
and its homotopy invariant, which is a 3-winding number/Wess-Zumino-Witten term 
\be\label{WZW}
	W_3[U_l]&=&\frac{1}{24\pi^2}\int_{T_3}d^2 \mathbf {k} dt ~
	\epsilon^{\mu\nu\rho} \nonumber\\
	&&\text{Tr} [(U_l^{\dagger}\partial_{\mu}U_l)(U_l^{\dagger}\partial_{\nu}U_l)(U_l^{\dagger}\partial_{\rho}U_l)]\,
\ee 
 were introduced to describe the quench dynamics for two-band Chern insulators. $\mathbf{h}$ and $\mathbf{h}_0$ are the flattened post-quench and pre-quench Hamiltonians with band-flattening given by the replacement
 $H\to \mathbf{h}=H/E_{\mathbf{k}}$ and $H_0\to \mathbf{h}_0=H_0/E_{0\mathbf{k}}$. Its value is equal to the difference of two static Chern numbers: 
\be \label{EqW3CfCi}
	W_3[U_l]=\mathcal{C}_f-\mathcal{C}_i\,,
\ee
in which $\mathcal{C}_{i/f}$ represent the Chern number of pre-quench and post-quench Hamiltonian.
Unlike the static Chern numbers in which band-flattening only serves as a computation technique and does not change the topology, the loop unitary and its homotopy invariant include band-flattening as part of the definition.  This is required to ensure the periodicity in the time direction. Otherwise the expression in Eq. (\ref{WZW}) does not give an integer number. This is the same situation in \cite{WangCe2017,ChenXin2020}.  The proof in their paper used Pauli matrices so it is limited to Clifford algebra. It is not clear from their paper whether their formalism is valid in the quench of generic multi-band Chern insulators, for instance for one with a Hamiltonian expanded by su(N) algebra. In the following, we show how the loop unitary is defined in such cases.
 
 For multi-band Chern insulators, we apply the band-flattening given by
\be\label{bandflat}
	H\to \mathbf{h}=\sum_a\sgn{(E_a)}P_a \quad
\mathbf{h}_0=\sum_a\sgn{(E_{0a})}P_{0a}\,,
\ee
in which $E_a $ and $E_{0a} $ are the energy eigenvalues, $P_a$ and $P_{0a}$ are the eigenprojectors and $a$ labels the energy bands. 
Here the band-flattening is more restricted: all the energy eigenvalue below the Fermi surface should be the same and the ones above the Fermi surface, too. Other less restricted choices may give integer number, but will not satisfy Eq. (\ref{EqW3CfCi}).
And the loop unitary and its homotopy invariant take the same form as in Eq. (\ref{LU}) and Eq. (\ref{WZW}) respectively. With Eq. (\ref{bandflat}) the multi-band loop unitary can be recast into 
\be
	U_l(t)=(\cos{t}-i\mathbf{h}\sin{t} )(\cos{t}+i\mathbf{h}_0\sin{t} )\,.
\ee 
One can check that $U_l(t+\pi)=U_l(t)$ so the $\pi$-periodicity of time is ensured. For a generic N-band Hamiltonian, one can check that the band-flattening Eq. (\ref{bandflat}) makes $U_l(t)$ an $U(N)$ group element,  but if the numbers of negative bands for post-quench and pre-quench Hamiltonian are the same then $U_l(t)$ is a $SU(N)$ group element. Both $\pi_3[U(N)]$ and $\pi_3[SU(N)]$ are integer valued. As time evolves, this $W_3[U_l]$ counts the winding of the mapping from $t,\mathbf{k}$ space to the generalized Bloch sphere for $U(N)$ groups\cite{JAKOBCZYK2001383,Zyczkowski2003,KIMURA2003339,PhysRevA.68.062322,PhysRevB.104.085114}.
The value of $W_3[U_l]$ is again equal to the difference of two static Chern numbers of bands below Fermi surfaces as in (\ref{EqW3CfCi}). If we connect the system with a chemical potential, we can change the Fermi surface. Then the band-flattening becomes
\be\label{bandflatmu}
	H'&=&H-\mu\to \mathbf{h}'=\sum_a\sgn{(E_a-\mu)}P_a\nonumber\\
	H'_0&=&H'_0-\mu\to \mathbf{h}'_0=\sum_a\sgn{(E_{0a}-\mu)}P_{0a}\,.
\ee
Since chemical potential term is proportional to identity matrix, the eigenfunctions are not changed at all. But the number of bands included in the static Chern numbers is changed. Therefore, in multi-band Chern insulators, there is a way to measure static Chern number of each band through quench dynamics.

In Appendix(\ref{W3CfCi}) we give the complete proof of Eq. (\ref{EqW3CfCi}) for a generic multi-band case using a different method  compared with \cite{HuHaiping2020}, which is down-to-earth algebraic. {The proof developed in \cite{HuHaiping2020} for two-band systems is made of two parts: firstly the 3-winding number of $U_l$ is equal to the difference of two Abelian Chern-Simons integrals and secondly these two Chern-Simons integrals reduce to Chern number after integration of time, which is given in \cite{WangCe2017}. For multi-band systems, finding the correct non-Abelian Chern-Simons integrals corresponding to the 3-winding number of $U_l$, and their relations with static Chern number are nontrivial problems. Here we omit the path to Chern-Simons action and instead we directly show that in the multi-band case that the 3-winding number is equal to static Chern number with the help of projection operators.} Our idea is very intuitive: first we show that $W_3[U_l]=W_3[e^{-i\mathbf{h}t}]+W_3[e^{-i\mathbf{h}_0t}]$ where $W_3[e^{-i\mathbf{h}t}]$ and $W_3[e^{-i\mathbf{h}_0t}]$ are obtained from replacing $U_l$ by $e^{-i\mathbf{h}t}$ and $e^{-i\mathbf{h}_0t}$ respectively(the result is mentioned in\cite{PhysRevB.96.195303}). Second we show $W_3[e^{-i\mathbf{h}t}]=\mathcal{C}_f$ and $e^{-i\mathbf{h}_0t}=-\mathcal{C}_i$. The first step is simple. 

Let us explain a little bit on the second step: {With the band-flattening Eq. (\ref{bandflat}) all the energy eigenvalues above(below) Fermi surface are degenerate so we can write 
\be
	e^{-i\mathbf{h}t}=\sum_a e^{iE_a t} P_a(\mathbf {k})=e^{iE t} P_++e^{-iE t} P_-\,,
\ee}
where in $P_+$ we include all the bands with energy above zero and  $P_-$ energy below zero 
and get
\be	
	W_3[e^{-i\mathbf{h}t}] &=&-\frac{i}{2\pi}\int_{T_2} \sum_{E_a<0}\text{Tr}P_adP_a \wedge dP_a \,,
\ee
which is the sum of Chern numbers of negative energy bands. Similar mathematical proof was also used in showing the relation between Chern number and the 3-winding number in Floquet systems \cite{PhysRevX.3.031005,PhysRevX.6.021013}.  If we choose other types of band-flattening than Eq. (\ref{bandflat}), such that the energies differ from each other, we will get 
\be	
	W_3[e^{-i\mathbf{h}t}] &=&-\frac{1}{4\pi^2}\text{Tr} \sum_a  \int_{T_3} d(iE_a t)\wedge P_adP_a\wedge dP_a\, \,
\ee
instead. Such $W_3[U_1]$ may not even be an integer because the periodicity may not be ensured and in general it is not proportional to the static Chern number. 
With the picture of gapless fermions given in the next section, we see that Eq. (\ref{EqW3CfCi}) relates two-dimensional gapped fermions with generic three-dimensional multifold gapless fermions.
\section{Phase band Chern number}\label{phasebandchern}
In this section we show that the 3-winding number manifest as gapless fermions in the phase bands of the loop unitary. In phase-band formalism 
\be\label{LUphase}
	U_l=\sum_a e^{i\phi_a(\mathbf {k},t)}|\phi_a\rangle \langle \phi_a| 
	=\sum_a e^{i\phi_a(\mathbf {k},t)}P_a(\mathbf {k},t)\,.
\ee
{For the simplicity of notation, we use $P_a$ in both Sec II and Sec III but the reader should be aware that different from the eigenprojector in Eq. (\ref{bandflat}), the  eigenprojectors in Eq. (\ref{LUphase}) mean the eigenstate of this loop unitary and this  depend on time. The notion of the phase band was introduced to describe the topology in Floquet systems and the physical meaning of the phase band remains to be explored. Likewise the physical meaning of the phase band in quench process is unknown either.} One should also beware the difference between usual energy bands and the phase bands: the gap-closing points are built-in structure of the periodic unitary time-evolution at $\sin{(\phi_a)}=0$ and are thus not necessarily related with the topology. However, among all the gap-closing points there exist fermions that do count the topological number. 

{For two-band case, the phase band representation involves only Pauli matrices, and as a result the spin vectors of two bands are just opposite to each other, which makes the computation and the topological structure relatively simple. In the multiband case, the algebra is much more complicated and the corresponding pseudo-spin vectors do not have simple relations between each other. However, we find that the contributions from different bands are decoupled from each other and finally the expression of the 3-winding number in terms of phase bands reduces to a simple form.} In Appendix(\ref{Phaseband}) we showed that
\be\label{W3PPP}
	W_3[U_l]=\frac{1}{4\pi^2}\text{Tr} \sum_a  \int_{T_3} d(i\phi_a)\wedge P_adP_a\wedge dP_a\,.
\ee
Eq. (\ref{W3PPP}) is a key formula of our work. This is a sum of topological numbers of each phase band. The periodicity  $\phi_a$ is required by that of $U_l$. Since every $\phi_a$ is periodic, the topological number of each phase band is an integer. Our proof is an algebraic proof, so Eq. (\ref{W3PPP}) applies to any system with a Hermitian Hamiltonian. This result is valid  and potentially useful for dynamical topological invariants even in integer quantum Hall effect.

The outline of the proof is as following:
Starting from
\be
	W_3[U_l]&=&\frac{1}{24\pi^2}\int_{T_3} 
	\text{Tr} [(U_l^{-1}d U_l)^3]
\ee
and using
\be
	U_l^{-1}d U_l&=&\sum_a e^{-i\phi_a}P_a d (\sum_b e^{i\phi_a}P_j)\nonumber
	\\
	&=& \sum_aP_ad(i\phi_a) +\sum_{a,b}e^{-i(\phi_a-\phi_b)}P_adP_b
\ee
\begin{widetext}
we have 
\be
\begin{aligned}
W_3[U_l]=&\frac{1}{24\pi^2}\int_{T_3} \text{Tr}\left(U_{l}^{\dagger} d U_{l}\right)^{3} \\
 =&\frac{1}{24\pi^2}\int_{T_3} \text{Tr}\Big(3 { \sum_{a} P_{a} d\left(i \phi_{a}\right) \sum_{b, c} e^{i\left(\phi_{0}-\phi_{c}\right)} P_{c} d P_{b} \sum_{e, f} e^{i\left(\phi_{e}-\phi_{f}\right)} P_{f} d P_{e}}\\
 & +{\sum e^{i\left(\phi_{a}-\phi_{b}\right)+i\left(\phi_{c}-\phi_{d}\right)+i\left(\phi_{e}-\phi_{f}\right)} P_{b} d P_{a} \cdot P_{d} d P_{c} P_{f} d P_{e}}\Big) \,.
\end{aligned}
\ee
\end{widetext}
The first and the second term are calculated separately as
\be
 	\text{1st term}&=& 2\sum_{a,b}d(i\phi_a) \wedge \text{Tr}(P_adP_a\wedge dP_a)\nonumber\\
	&&-\sum_{a,b}de^{-i(\phi_a-\phi_b)} \wedge \text{Tr}(P_adP_a\wedge dP_b)\,.\\
	\text{2nd term}&=&-3 \sum_{f, e} e^{-i\left(\phi_f-\phi_e\right)}  d \operatorname{Tr}\left(P_f \wedge d P_f \wedge d P_e\right)\,.
\ee
Summing over the two terms we get
\be
	\textcircled{1}+\textcircled{2}&=&6\text{Tr}  \sum_a  d(i\phi_a)\wedge P_adP_a\wedge dP_a\nonumber \\
	&&-3\sum_{a,b}d(e^{-i(\phi_a-\phi_b)} \text{Tr}(P_a dP_a \wedge d P_b))
\ee
and the second line vanishes after integration in $T^3$. Therefore, we get Eq. (\ref{W3PPP}).

Using the formula 
\be
	\nabla \phi_a=\pi\sgn{(\cos{\phi_a})}\delta (\sin{\phi_a}) \nabla \sin{\phi_a}
\ee
from \cite{PhysRevLett.127.196404}
we rewrite Eq. (\ref{W3PPP}) as
\be\label{W3Ome}
	W_3[U_l]=\frac{1}{4\pi}\sum_a\sgn{(\cos{\phi_a})}  \int_{\mathbf{S}^a_p}d \mathbf{S}^a_p\cdot \mathbf{\Omega}^a  \Big|_{\sin{\phi_a}=0}\,.
\ee
\be\label{Ome}
\mathbf{\Omega}^a=i\text{Tr}(P_a   \nabla P_a\times  \nabla P_a)
\ee
is the Berry curvature of $U_l$ and 
$\mathbf{S}^a_p$ is in the direction of  $\nabla \sin{\phi_a}$,  surrounding the band-crossing points in the phase band of $\phi_a$, and this gives opposite sign for phase bands with increasing $\phi_a$ and decreasing $\phi_a$ during time evolution. Eq. (\ref{W3Ome}) gives another interpretation of $W_3[U_l]$, which is topological number of gapless fermions of phase bands. The way to evaluate this topological number is as following:  We solve the eigenvector or the eigenprojector of $U_l$, calculate the Berry curvature and integrate it over $(\mathbf {k}, t)$ space around the gapless points and then take a sum over all the points. The details will be given in Sec. (\ref{Mutilfoldfermion}).

For an su(N) Hamiltonian expanded by generalized Gell-Mann matrices $\lambda^m$ the coherence vector is defined as
\be
	S_a^m:=\text{Tr}(\lambda^mP_a)\Leftrightarrow P_a=:\frac{\mathbb{1}}{N}+\frac{1}{2}S_a^m \lambda^m \,.
\ee 
In such cases we can rewrite Eq. (\ref{W3Ome}) as 
\be\label{W3USSS}
	W_3[U_l]
	&=&  \frac{1}{16\pi}\sum_a\sgn{(\cos{\phi_a})}\nonumber\\ &&\int_{\mathbf{S}^a_p}d \mathbf{S}^a_p\cdot  f_{mnl} (S_a^m \nabla S_a^n \times \nabla S_a^l)\Big|_{\sin{\phi_a}=0}\,,
\ee
where $f_{mnl}$ is the anti-symmetric structure constant for su(N) algebra.(See Appendix(\ref{Phaseband}) for derivation.)  We see that besides the $\pi$-defect, the defect on $\phi_a=0$, which we call 0-defect, may appear to represent the topological charge as well.  It is easy to see that for 2-by-2 matrix this reduces to the formula for $\pi$-defect in \cite{HuHaiping2020}. 

\section{Multifold fermions in quench dynamics}\label{Mutilfoldfermion}
The band-crossing points appearing in the quench dynamics of generic multi-band Chern insulators can take the form of not only Weyl fermions but also multifold fermions\cite{BradlynScience2016}, which cannot happen in two-band quench process. In this section, we show an example how a three-fold fermion appears in the phase band in the quench process.

Let us first explain the procedure of identifying a chiral gapless fermion which contributes to the dynamical winding number. {Eq. (\ref{W3Ome}) tells us that contributions come from the gapless points at $\phi_a=0 $ or $\phi_a=\pi$ , which means 
\be
	\det (U_l\pm \mathbf{1})=0\,.
\ee 
We choose appropriate points $(t,k_x, k_y)=(\bar{t},\bar{k}_x, \bar{k}_y)$ to ensure this requirement, 
 then expand $U_l$ near these points.
 \be
 	U_l(t,k_x, k_y)=U_l(\bar{t},\bar{k}_x, \bar{k}_y)+U_l(\delta t, \delta k_x, \delta k_y)\,.
 \ee
 To get the value of dynamical winding number, we do the following
 \\
(1) Solve the eigen problem
$$
U_l( t,  k_x,  k_y)\left|\phi_a\right\rangle=e^{i \phi_a}\left|\phi_a\right\rangle ;
$$
(2)Calculate the topological charge $$\frac{1}{4\pi}  \int_{\mathbf{S}^a_p}d \mathbf{S}^a_p\cdot \mathbf{\Omega}^a  \Big|_{\sin{\phi_a}=0}$$ around $(\bar{t},\bar{k}_x, \bar{k}_y)$ with 
Berry curvature given by Eq. (\ref{Ome}). A nontrivial topological number corresponds to a chiral gapless fermion.
}

Next we show an explicit example. We choose the pre-quench Hamiltonian as  
\be
	H_0=\lambda_7\,,
\ee
and guided by intuition,
the post-quench Hamiltonian as
\be
\begin{aligned}
H & =h_x \lambda_2+h_y \lambda_5+h_z \lambda_7 \\
& =\left[\begin{array}{ccc}
0 & -i h_x & -i h_y \\
i h_x & 0 & -i h_z \\
i h_y & i h_z& 0
\end{array}\right]
\end{aligned}
\ee 
with coefficients $h_x=\sin k_x\,,~h_y=\sin k_y\,,~h_z=M-\cos k_x -\cos  k_y$ and 
 eigenvalues
\be
 E_i= \pm h, 0= \pm \sqrt{\sum_{i=1}^3 h_i^2}, 0\,.
\ee
The post-quench Hamiltonian $H$ looks similar as the two Qi-Wu-Zhang model but there are two key differences: First, there is a flat-band with zero Berry curvature between two other bands; second the Chern numbers of the two other bands are twice as that in Qi-Wu-Zhang model with same coefficients $h_i$. Strictly speaking, this is a conductor. (We can shift the total energy to make it an insulator. {However, the gapless point is shifted from the flat phase band.} See Appendix (\ref{Phasebandenergyshift})). By band-flattening we have
\be
\begin{aligned}
\mathbf{h} & =(h_x \lambda_2+h_y \lambda_5+h_z \lambda_7)/h=\hat{h}_x \lambda_2+\hat{h}_y \lambda_5+\hat{h}_z \lambda_7 \\
\mathbf{h}_0 & =\lambda_7
\end{aligned}
\ee
The loop unitary in this case is
 \begin{align}
U_l  &=e^{i \mathbf{h} t} e^{-i \mathbf{h}_0 t}\nonumber \\
& =\left(\cos t ~\mathbf{h}^2+i \sin t ~\mathbf{h}+P_0\right) \cdot\left(\cos t ~\lambda_7^2-i \sin t ~\lambda_7+P_0^0\right)\,,\nonumber\\
\end{align}
in which $P_0$ and $P_0^0$ are the projection operators of the zero-energy bands of post- and pre-quench Hamiltonians respectively. A unique point of this example is that the zero energy band breaks the $\pi$-periodicity of time into $2\pi$-periodicity.
\begin{figure}[h]
	\centering  %
	\includegraphics[width=1\linewidth]{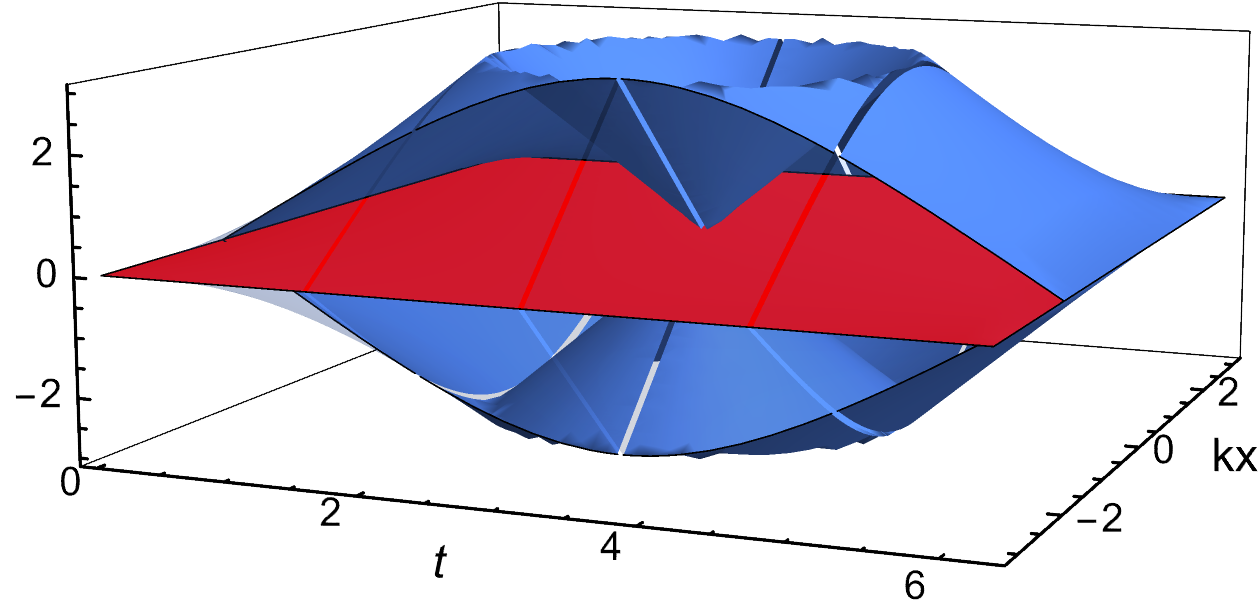} 	\caption{Three phase band dispersions at $k_y=0$, we made the bands transparent for the region $-\pi<k_x<-\pi/2$.}  %
	\label{phasebanddispersion}   %
\end{figure}
From Fig.(\ref{phasebanddispersion}) we see there is a gapless point at $t=\pi, k_x=0$. We call it a 0-defect because $U_l(t=\pi,k_x=k_y=0)=\mathbb{1}$ meaning the phase is 0 at this gapless point.
We expand the loop unitary near this point:
{\be\label{Uldeltat}
	U_l&=&e^{i \mathbf{h}(\delta t+\pi)} e^{-i \mathbf{h}_0(\delta t+\pi)}
\nonumber\\
& \doteq& \mathbb{1}-i \sgn{(M-2)}( \delta k_x\lambda_5-\delta k_y \lambda_2  )\nonumber
\\
&&+i \left(\sgn{(M-2)}-1\right)\delta t \lambda_7 \,.
\ee}
The detailed calculation is in Appendix(\ref{Expansion}).
From the form it is easy to tell that this is a three-fold fermion mentioned in \cite{BradlynScience2016} carrying topological number 2. 

\begin{figure}[h]
	\centering  %
	\includegraphics[width=0.8\linewidth]{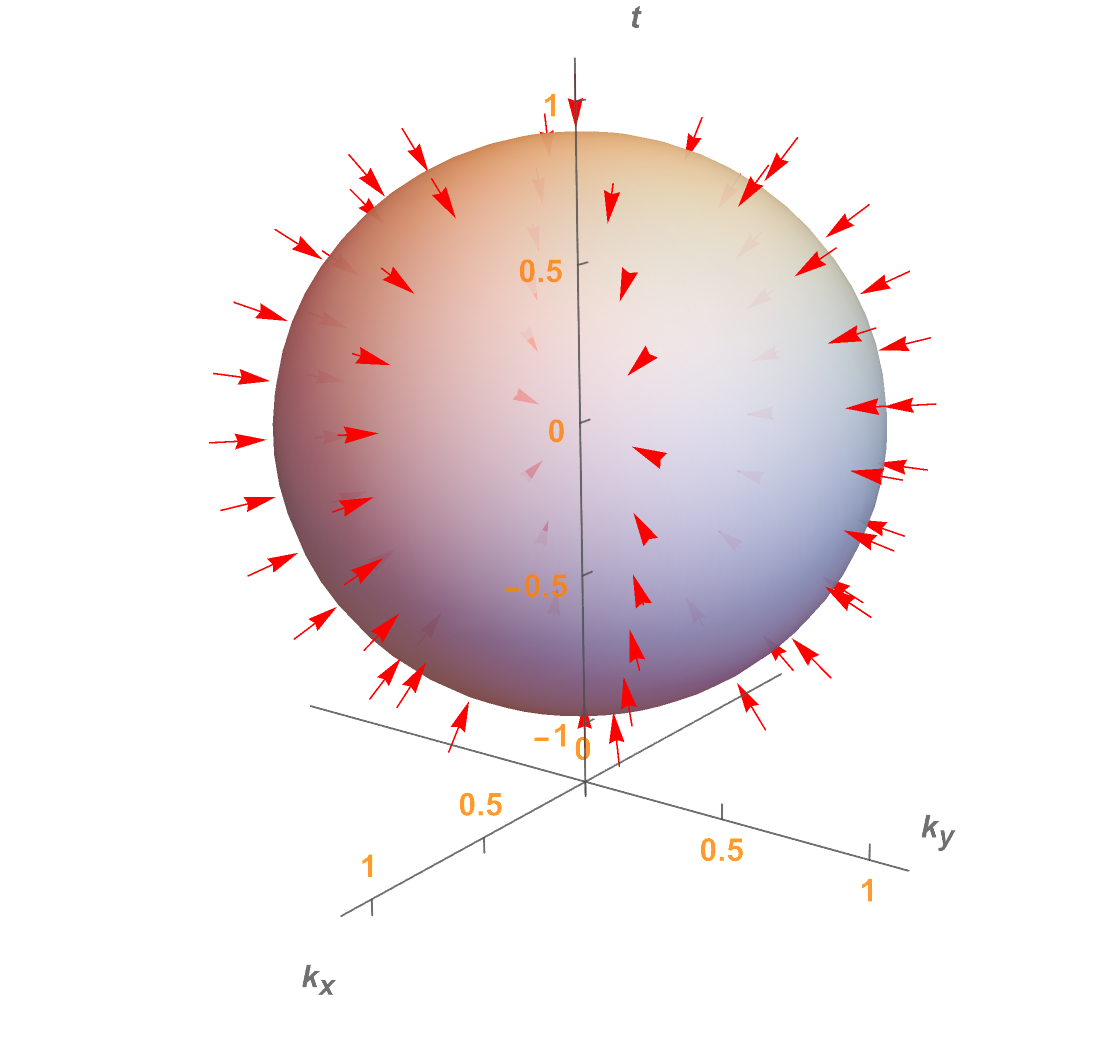}  %
	\caption{Berry curvature vector of the lowest phase band of the 0-defect}  %
	\label{Berrycur}   %
\end{figure}

From Eq. (\ref(\ref{W3USSS}) we see that the coherence vector $S^m_a$ have $N^2-1$ components for each band for su(N), so it is no longer practical to draw them as a vector field when $N>2$. However, the Berry curvature always have three components.
For the convenience of visualization, we draw the field of Berry curvature to represent the defects. Fig. (\ref{Berrycur}) shows the Berry curvature of 0-defect we got from the example in this section.

\section{Summary and Discussion}\label{SumDisc}
In this work, we studied the quench dynamics of multi-band Chern insulators in two-dimensions. We showed that for any number of bands, the difference of static Chern numbers of post-quench and pre-quench Hamiltonian can be represented by topological gapless fermions in the phase bands of the loop unitary operator. Moreover, we provide an example that a three-fold gapless fermions arises in the phase bands in a quench of a three band model. We expect that gapless fermions with multi-fold degeneracy higher than three may appear if the Chern insulator has more  bands. Further on this direction, one wonders whether other types of topological gapless structures such as nodal lines can appear in the phase bands and represent the static topology during quench dynamics. We leave this as a future exploration.

{
Both in the case of two-band models in\cite{HuHaiping2020} and our three-band models,  we have checked that there is only one chiral fermion in the whole period in the $(\mathbf{k},t)$ space.} It is not surprising from the construction of our theory because the total chirality represents the difference of the static Chern numbers which must be nonzero in nontrivial quenches. This is also consistent with the results in \cite{PhysRevB.82.235114,PhysRevLett.123.066403} that a single Weyl fermion exists in the quasienergy band of a Floquet unitary operator. The nature of phase band is different from energy band such that it does not satisfy the traditional Nielsen-Ninomiya theorem\cite{NIELSEN198120,NIELSEN1981173}.  

\cite{PhysRevResearch.4.023120} shows the protocol to mapped out all the coherence vectors through time-of-flight images in the generalized Bloch sphere for su(N) Hamiltonians. \cite{PhysRevLett.122.253601} gives a method to measure the defects of phase bands directly for two band models, which can in principle be generalized. The combination of both methods may help detect defects in quench of multi-band models.

{As explained in \cite{HuHaiping2020}, the Hopf invariant defined in \cite{WangCe2017,ChenXin2020} works only for the case when the prequench Hamiltonian is topologically trivial and it suffers from the lack of gauge invariance. Suppose one accept these and define the Chern-Simons invariant for multi-band Hamiltonian quench processes, there is another challenge that gives limitation on the practical level: The Chern-Simons invariant becomes non-Abelian and the  one dimensional  linking number is replaced by at least a three dimensional one\cite{Si2005} where the dimension increases as the number of bands below Fermi surfaces grows so drawing the links becomes more and more complicated.}
However, the Berry curvature vector of the loop unitary are always three dimensional for the quench of two dimensional Chern insulators so it is rather convenient.

\section*{Acknowledgements}
This work was supported by the National Natural Science Foundation of China (Grants No. 11905054 and No. 12275075) and Doctoral scientific research foundation of Henan Normal University(Grants No. 5101029170913).

\appendix
\section{The proof that $W_3=\mathcal{C}_f-\mathcal{C}_i$}\label{W3CfCi}
We start from 
\be
	W_3[U_l]&=&\frac{1}{24\pi^2}\int_{T_3}d^2 \mathbf {k} dt ~
	\epsilon^{\mu\nu\rho} \nonumber\\
	&&\text{Tr} [(U_l^{-1}\partial_{\mu}U_l)(U_l^{-1}\partial_{\nu}U_l)(U_l^{-1}\partial_{\rho}U_l)]
	\\
	&=&\frac{1}{24\pi^2}\int_{T_3} 
	\text{Tr} [(U_l^{-1}d U_l)^3]\,.\label{WZWd3}
\ee
We prove $W_3=\mathcal{C}_f-\mathcal{C}_i$ in two steps: First, since $U_l=e^{-iht}e^{ih_0t}=U_1U_2$ where $U_1=e^{-iht}, U_2=e^{ih_0t}$, we show that 
\be\label{WlW1W2}
	W_3[U_l]=W_3[U_1]+W_3[U_2]
\ee	
where
 \be
 	 W_3[U_a]&=&\frac{1}{24\pi^2}\int_{T_3}d^2 \mathbf {k} dt ~
	\epsilon^{\mu\nu\rho} \nonumber\\
	&& \text{Tr} [(U_a^{-1}\partial_{\mu}U_a)(U_a^{-1}\partial_{\nu}U_a)(U_a^{-1}\partial_{\rho}U_a)]\,, a=1,2\nonumber\,.
\ee
	Second
\be
	W_3[U_1]=\mathcal{C}_f \label{WCf}
	\\
	W_3[U_2]=-\mathcal{C}_i\label{WCi}\,.
\ee

Let's show Eq. (\ref{WlW1W2}) in the following.
\begin{widetext}
\be
	\text{Tr}  [(U_l^{-1}d U_l)^3]&=&\text{Tr}  [(U_2^{-1}U_1^{-1}d (U_1U_2))^3]\nonumber
	\\
	&=&\text{Tr}  [(U_2^{-1}d U_2+U_2^{-1}U_1^{-1}d U_1U_2)^3]\nonumber
	\\
	&=&\text{Tr}  [(U_2^{-1}d U_2)^3+(U_1^{-1}d U_1)^3+3dU_2\wedge U_2^{-1}dU_2\wedge U_2^{-1}U_1^{-1} d U_1+3U_2^{-1}U_1^{-1}dU_1 U_1^{-1}dU_1]\nonumber
	\\
	&=&\text{Tr}  [(U_2^{-1}d U_2)^3+(U_1^{-1}d U_1)^3-3dU_2\wedge dU_2^{-1}\wedge U_1^{-1} d U_1-3dU_1^{-1} \wedge dU_1 \wedge dU_2 U_2^{-1}]\nonumber
	\\
	&=&\text{Tr}  [(U_2^{-1}d U_2)^3+(U_1^{-1}d U_1)^3-3U_1^{-1} d U_1 \wedge  d(U_2\wedge dU_2^{-1})+3d(U_1^{-1} \wedge dU_1) \wedge U_2 d U_2^{-1}]\nonumber
	\\
	&=&\text{Tr}  [(U_2^{-1}d U_2)^3+(U_1^{-1}d U_1)^3+3d(U_1^{-1} d U_1 \wedge  U_2 dU_2^{-1})]\,.
\ee
\end{widetext}
Since $U_1$ and $U_2$ are periodic in $\mathbf{k}$ and $t$, the last term vanishes after integration and we get Eq. (\ref{WlW1W2}). Moreover, 
\be
	\text{Tr}(U_2^{-1}d U_2)^3 = -\text{Tr}(dU_2^{-1} U_2)^3=-\text{Tr}(U_2dU_2^{-1} )^3
\ee
so Eq. (\ref{WlW1W2}) can be rewritten as
\be
	W_3[U_l]=W_3[U_1]-W_3[U_2^{-1}]\,.
\ee
Next let's show Eq. (\ref{WCf}) and (\ref{WCi}), and this is an appetizer for a more general result in the next section. We start with 
\be
	U_1=\sum_a e^{-iE_a t} P_a(\mathbf {k})=e^{-iE t} P_++e^{iE t} P_-\,,
\ee
where in $P_+$ we include all the bands with energy above zero and  $P_-$ energy below zero and we assume band-flattening for all the bands(this assumption is removed in the proof in Appendix \ref{Phaseband}). Then 
\begin{widetext}
\be\label{W3dtdPdP}
	W_3[U_1]&=&\frac{1}{24\pi^2}\int_{T_3} 
	\text{Tr} [(U_1^{-1}d U_1)^3]\nonumber
	\\
	&=&\frac{1}{24\pi^2}\int_{T_3} 
	\text{Tr} ((e^{iE t} P_++e^{-iE t} P_-)d (e^{-iE t} P_++e^{iE t} P_-))^3\nonumber
	\\
&=&\frac{1}{24\pi^2}\int_{T_3} 
	\text{Tr} \Big(-(P_+-P_-)d(iEt)+((e^{2iEt}-1)P_+-(e^{-2iEt}-1)P_-)dP_-\Big)^3\nonumber\\
	&=&\frac{-1}{8\pi^2}\int_{T_3} 
	\text{Tr} \Big[(P_+-P_-)d(iEt)\Big(((e^{2iEt}-1)P_+-(e^{-2iEt}-1)P_-)dP_-\Big)^2\Big]
\ee
in the last step we used the fact that only the terms proportional to $d t$ remains and $P_-$ is independent of $t$. Then we expand Eq. (\ref{W3dtdPdP}) using
\be
	\text{Tr}(P_+ d P_+ \wedge P_+ d P_+)=\text{Tr}(P_- d P_- \wedge P_- d P_-)=0
\ee
to get
\be	
	W_3[U_1]&=&\frac{-1}{4\pi^2}\int_{T_3} 
	\text{Tr} \Big[d(iEt) \wedge ((1-e^{2iEt})(e^{-2iEt}-1))P_+dP_- \wedge P_-dP_- \Big]\nonumber\\
	&=&\frac{-1}{4\pi^2}\int_{T_3} 
	 d(iEt) \wedge ((2-\cos{(2Et)})\text{Tr}P_-dP_- \wedge dP_-\nonumber\\
	 &=&\frac{-1}{2\pi^2}\int_{T_3} 
	 d(iEt) \wedge \text{Tr}P_-dP_- \wedge dP_- \nonumber\\
	  &=&\frac{-i}{2\pi}\int_{T_2} \text{Tr}P_-dP_- \wedge dP_-\nonumber\\
	  &=&\frac{-i}{2\pi}\int_{T_2} \sum_{E_a<0}\text{Tr}P_adP_a \wedge dP_a \,.\label{W3U1}
\ee
\end{widetext}

Therefore, we see from Eq. (\ref{W3U1}) that $W_3[U_1]$ is indeed the static Chern number. And calculation for $W_3[U_2]$ is the same so we get the conclusion of Eq. (\ref{WCf}) and (\ref{WCi}). Thus $W_3=\mathcal{C}_f-\mathcal{C}_i$ is proved.

\section{The proof of Eq. (\ref{W3PPP})}\label{Phaseband}
In this section, we solve $W_3$ to get the phase band topological number. 
We again use Eq. (\ref{WZWd3})
and use $U_l=\sum_a e^{i\phi_a(\mathbf {k},t)}P_a(\mathbf {k},t)$ where $e^{i\phi_a(\mathbf {k},t)}$ are the phase bands and $P_a(\mathbf {k},t)$ the corresponding projection operators, to expand it. We generalize the calculation of Eq. (\ref{W3U1}) and prove the identity
\be\label{W3PPPB}
	W_3[U_l]=\frac{1}{4\pi^2}\int_{T_3}\text{Tr} \sum_a  d(i\phi_a)\wedge P_adP_a\wedge dP_a\,.
\ee
 Note that
\be
	U_l^{-1}d U_l&=&\sum_a e^{-i\phi_a}P_a d (\sum_b e^{i\phi_a}P_j)\nonumber
	\\
	&=& \sum_aP_ad(i\phi_a) +\sum_{a,b}e^{-i(\phi_a-\phi_b)}P_adP_b
\ee
and the integrant in Eq. (\ref{WZWd3}) can be written as 
\begin{widetext}
\be
\begin{aligned}
& \text{Tr}\left(U_{l}^{\dagger} d U_{l}\right)^{3} \\
 =&\text{Tr}\left(\sum_{a} P_{a} d\left(i \phi_{a}\right)+\sum_{a, b} e^{i\left(\phi_{a}-p_{b}\right)} P_{b} d P_{a}\right)^{3} \\
=&  \text{Tr}\left(\left(\sum P_a d\left(i \phi_a\right)\right)^3+3(\sum_a P_a d\left(i \phi_a\right))^2 \sum_{c, d} e^{i\left(\phi_c-\phi_b\right)} P_d d P_c\right. \\
& +3 \sum_a P_a d\left(i \phi_a\right)(\sum_{b, c} e^{i\left(\phi_{b^{-}} \phi_c\right)} P_c d P_b)^2 \left.+(\sum_{a, b} e^{i\left(\phi_{a}-\phi_{b}\right)} P_{b} d P_{a})^{3}\right) \\
 =&\text{Tr}\Big(3 \underbrace{ \sum_{a} P_{a} d\left(i \phi_{a}\right) \sum_{b, c} e^{i\left(\phi_{0}-\phi_{c}\right)} P_{c} d P_{b} \sum_{e, f} e^{i\left(\phi_{e}-\phi_{f}\right)} P_{f} d P_{e}}_{\textcircled{1}} \\
& +\underbrace{\sum e^{i\left(\phi_{a}-\phi_{b}\right)+i\left(\phi_{c}-\phi_{d}\right)+i\left(\phi_{e}-\phi_{f}\right)} P_{b} d P_{a} \cdot P_{d} d P_{c} P_{f} d P_{e}}_{\textcircled{2}}\Big) 
\end{aligned}
\ee

Using 
\be
	\begin{aligned}
& P_{b} d P_{a}=d\left(P_{b} \delta_{a b}\right)-d P_{b} P_{a}=d P_{b}\left(\delta_{a b}-P_{a}\right) \\
& P_{b} d P_{a} P_{d}=d P_{b}\left(\delta_{a b} P_{d}-\delta_{a d} P_{d}\right)=\left(\delta_{a b}-\delta_{a d}\right) d P_{b} P_{d}
	\end{aligned}
\ee
we calculate the two terms in the integrants separately.

\be
	\begin{aligned}
\textcircled{1}&= \operatorname{Tr}\left(\sum_{a} P_{a} d\left(i \phi_{a}\right) \sum_{b, c} e^{i\left(\phi_{b}-\phi_{c}\right)} P_{c} d P_{b} \sum_{e, f} e^{i\left(\phi_{e}-\phi_{f}\right)} P_{f} d P_{e}\right) \\
& =\operatorname{Tr} \sum d\left(i \phi_{a}\right) e^{i\left(\phi_{b}-\phi_{a}\right)} P_{a }d P_{b} e^{i\left(\phi_{e}-\phi_{f}\right)} P_{f} d P_{e} \\
& =\operatorname{Tr} \sum d\left(i \phi_{a}\right) e^{i\left(\phi_{b}-\phi_{a}\right)+i\left(\phi_{e}-\phi_{f}\right)} P_{a} d P_{b} P_{f} d P_{e} \\
& =\operatorname{Tr} \sum d\left(i \phi_{a}\right) e^{i\left(\phi_{b}-\phi_{a}\right)+i\left(\phi_{e}-\phi_{f}\right)}\left(\delta_{a} b-\delta_{b f}\right) d P_{a} P_{f} d P_{e} \\
& =\operatorname{Tr} \sum\left(d\left(i \phi_b\right) e^{i\left(\phi_e-\phi_f\right)} d P_b P_f d P_e +d\left(i \phi_a\right) e^{i\left(\phi_e-\phi_a\right)} d P_a P_f d P_e\right) \\
& =-\operatorname{Tr} \sum d\left(i \phi_{b}\right) e^{i\left(\phi_{e}-\phi_{f}\right)} P_{f} d P_{e} d P_{b}
\end{aligned}
\ee	
\end{widetext}

  \begin{widetext}
 In order to proceed, we use
 \be
 	&&\text{Tr}(P_a dP_b \wedge d P_c)\nonumber\\
	&=&\langle \phi_a| d(|\phi_b\rangle\langle \phi_b|) \wedge d(|\phi_c\rangle\langle \phi_c|) |a \rangle \nonumber
	\\
	&=&\langle \phi_a| d|\phi_b \rangle \langle \phi_b| \wedge d|\phi_c \rangle \delta_{ca} + \langle \phi_a| d|\phi_b \rangle d \langle \phi_c|  |a \rangle\delta_{bc} 
	\\
	&&+\delta_{ab}\delta_{ca}d \langle \phi_b| d| c \rangle +\delta_{ab}d\langle \phi_b||\phi_c \rangle d\langle \phi_c||\phi_a \rangle \nonumber
	\\
	&=&\delta_{ca}\text{Tr}(P_a dP_a \wedge d P_b)+\delta_{bc} \text{Tr}(P_a dP_b \wedge d P_b)\nonumber
	\\
	&&+\delta_{ab}\delta_{ca} \text{Tr}(P_a dP_a \wedge d P_a)+\delta_{ab} \text{Tr}(P_a dP_a \wedge d P_c)\,.
\ee
 Then the first term becomes
 \be
 	\textcircled{1}&=& 2\sum_{a,b}d(i\phi_a) \wedge \text{Tr}(P_adP_a\wedge dP_a)\nonumber\\
	&&-\sum_{a,b}de^{-i(\phi_a-\phi_b)} \wedge \text{Tr}(P_adP_a\wedge dP_b)\,.
 \ee

 Next let us calculate the second term. Using

\be
\begin{aligned}
	& \operatorname{Tr}\left(P_{b} d P_{a} P_{d} d P_{c} P_{f} d P_{e}\right) \\
	= & \operatorname{Tr}\left(\left(\delta_{a b}-\delta_{a d}\right) d P_{b} P_{d} d P_{c} P_{f} d P_{e}\right) \\
	= & \operatorname{Tr}\left(\left(\delta_{a b}-\delta_{a d}\right)\left(\delta_{c d}-\delta_{c f}\right) d P_{b} d P_{d} P_{f} d P_{e}\right)
\end{aligned}
\ee
and 
\be
\begin{aligned}
	& \sum_{a, b, c, d, e, f} e^{i\left(\phi_{a}-\phi_{b}\right)+i\left(\phi_{c}-\phi_{d}\right)+i\left(\phi_{e}-\phi_{f}\right)}\left(\delta_{a b}-\delta_{a d}\right)\left(\delta_{c d}-\delta_{c f}\right) \\
	= & \sum\left(e^{i\left(\phi_{c}-\phi_{d}\right)+i\left(\phi_{e}-\phi_{f}\right)}-e^{i\left(\phi_{c}-\phi_{b}\right)+\left(\phi_{e}-\phi_{f}\right)}\right)\left(\delta_{c d}-\delta_{c f}\right) \\
	= & \sum\left(e^{i\left(\phi_{e}-\phi_{f}\right)}-e^{i\left(\phi_{d}-\phi_{b}\right)+i\left(\phi_{e}-\phi_{f}\right)}-e^{i\left(\phi_{e}-\phi_{d}\right)}+e^{i\left(\phi_{e}-\phi_{b}\right)}\right)
\end{aligned}
\ee
the second term becomes
\be
\begin{aligned}
	\textcircled{2}= & \sum (e^{i\left(\phi_{e}-\phi_{f}\right)}-e^{i\left(\phi_{d}-\phi_{b}\right)+i\left(\phi_{e}-\phi_{f}\right)}-e^{i\left(\phi_{e}-\phi_{d}\right)}
	\\ +e^{i\left(\phi_{e}-\phi_{b}\right)}) 
	&\operatorname{Tr} d P_{b} d P_{d} P_{f} d P_{e}\,.
\end{aligned}
\ee
Now for the four exponential terms, only the second one survives after integration:
\be\label{2febd}
\begin{aligned}
\textcircled{2} & =-\sum_{f, e, b, d} e^{-i\left(\phi_f-\phi_e\right)-i\left(\phi_b-\phi_d\right)} \operatorname{Tr}\left(P_f d P_e d P_b d P_d\right)
\end{aligned}
\ee
because
\be
\begin{aligned}
& \sum_{b,d, e, f} \operatorname{Tr}\left(e^{i\left(\phi_{e}-\phi_{f}\right)} d P_{b} d P_{d} P_{f} d P_{e}\right) \nonumber\\
 =&\sum \operatorname{Tr}\left(e^{i\left(\phi_{e}-\phi_{f}\right)} d P_{b}\left(\delta_{d f} d P_{f}-P_{d} d P_{f}\right) d P_{e}\right) \\
 =&\sum \operatorname{Tr}\left(e^{i\left(\phi_{e}-\phi_{f}\right)}\left(d P_{b} d P_{f} d P_{e}-d P_{b} P_{d} d P_{f} d P_{e}\right)\right) \\
 =&0
 \end{aligned}
\ee	
 \be
\begin{aligned}
& \int_{T^3} {\sum} \operatorname{Tr} e^{i\left(\phi_{e}-\phi_{d}\right)} d P_{b} d P_{d} P_{f} d P_{e} \\
=&\int_{T^3} e^{i\left(\phi_{e}-\phi_{d}\right)} d P_{b} d P_{d} d P_{e} \\
 =&\sum \operatorname{Tr} \int_{T^3}\left(d\left(e^{i\left(\phi_{e}-\phi_{d}\right)} P_{b} d P_{d} d P_{e}\right)-d e^{i\left(\phi_{e}-\phi_{d}\right)} P_{b} d P_{d} d P_{e}\right) \\
 =&\sum \operatorname{Tr} \int_{\partial T^3}\left(e^{i\left(\phi_{e}-\phi_{d}\right)} P_{b} d P_{d} d P_{e}\right) \\
=&0
\end{aligned}
\ee	
and with the same logic we have $$ \int_{T^3} {\sum} \operatorname{Tr} e^{i\left(\phi_{e}-\phi_{d}\right)} d P_{b} d P_{d} P_{f} d P_{e}=0\,.$$
Substituting into Eq. (\ref{2febd}) the following expression, which can be checked
\be
\begin{aligned}
& \operatorname{Tr}\left(P_f d P_e d P_b d P_d\right) \\
= & \left\langle\phi_f|d| \phi_e\right\rangle d\left\langle\phi_e|d| \phi_f\right\rangle \delta_{e b} \delta_{f d} +\left\langle\phi_f|d| \phi_b\right\rangle d\left\langle\phi_b|\cdot| \phi_f\right\rangle \delta_{f e} \delta_{b d} \\
+&d\left(\left\langle\phi_f|d| \phi_b\right\rangle\left\langle\phi_b|d| \phi_f\right\rangle\right) \delta_{f e} \delta_{d f} +\left\langle\phi_f|d| \phi_e\right\rangle\left\langle\phi_e|d| \phi_b\right\rangle\left\langle\phi_b|d| \phi_f\right\rangle \delta_{f d} \nonumber\\
+& \left\langle\phi_f|d| \phi_e\right\rangle\left\langle\phi_e|d| \phi_b\right\rangle d\left\langle\phi_b|\cdot| \phi_f\right\rangle \delta_{b d} 
+\left\langle\phi_f|d| \phi_e\right\rangle d\left\langle\phi_e|\cdot| \phi_d\right\rangle d\left\langle\phi_d|\cdot| \phi_f\right\rangle \delta_{e b} \\
 +&d\left\langle\phi_b|\cdot| \phi_d\right\rangle d\left\langle\phi_d|\cdot| \phi_f\right\rangle d\left\langle\phi_f|\cdot| \phi_b\right\rangle \delta_{f e} 
\end{aligned}
\ee
\end{widetext}
we get 
\be
 \textcircled{2}=-3 \sum_{f, e} e^{-i\left(\phi_f-\phi_e\right)} d \operatorname{Tr}\left(P_f d P_f d P_e\right)\,.
\ee

Summing over the two terms we get
\be
	\textcircled{1}+\textcircled{2}&=&6\text{Tr}  \sum_a  d(i\phi_a)\wedge P_adP_a\wedge dP_a\nonumber\\
	&-&3\sum_{a,b}d(e^{-i(\phi_a-\phi_b)} \text{Tr}(P_a dP_a \wedge d P_b))
\ee
and the second line vanishes after integration in $T^3$. Therefore, we get Eq. (\ref{W3PPPB}).

Now we use the projection operator formula for su(N) algebra
\be
	P_a=\frac{ \mathbb{1}}{N}+\frac{1}{2}\overrightarrow{S}_a\cdot \overrightarrow{\lambda}\,,~f_{mnl}=-\frac{i}{4}\text{Tr}(\lambda_m[\lambda_n,\lambda_l])
\ee
to rewrite Eq. (\ref{Ome}) as
\be
\mathbf{\Omega}^a_i 
& =&i \epsilon_{i j k} \operatorname{Tr}\left(P_a \partial_j P_a \partial_k P_a\right) \nonumber\\
& =&\frac{i}{8} \epsilon_{i j k} \operatorname{Tr}\left(\lambda_m \lambda_n \lambda_l\right) S_a^m \partial_j S_a^n \partial_k S_a^l \nonumber \\
& =&\frac{i}{16} \epsilon_{i j k} \operatorname{Tr}\left(\lambda_m\left[\lambda_n, \lambda_l\right]\right) S_a^m \partial_j S_a^n \partial_k S_a^l \nonumber
\\
& =&-\frac{1}{4} f_{m n l} S_a^m\left(\nabla S_a^n \times \nabla S_a^l\right)_i
\ee
and therefore Eq. (\ref{W3USSS}) is obtained.

\section{Expansion of $U_l$}\label{Expansion}
In this section, we explain the expansion of $U_l$  in Eq. (\ref{Uldeltat}) With the condition $h_x=\sin k_x\,,~h_y=\sin k_y\,,~h_z=M-\cos k_x -\cos  k_y$.
\be
&& \left.U_l\left(\delta t, \delta k_x, \delta k_y\right)\right|_{\left(t, k_x, k_y\right)=(\pi, 0,0)} \nonumber \\
 &=&\left(\cos t ~\mathbf{h}^2+i \sin t~ \mathbf{h}+P_0\right)\nonumber  \\
 &&\cdot\left(\cos t \lambda_7^2-i \sin t \lambda_7+P_0^0\right)\Big|_{\left(t, k_x, k_y\right) =(\pi, 0,0)}\nonumber  \\
& =&\left(-\left(\lambda_7^2+\delta \mathbf{h}^2\right)+P_0^0+i(-\delta t) \hat{h}_z \lambda_7\right)\nonumber \\
&& \cdot\left(-\lambda_7^2+P_0^0+i \delta t \lambda_7\right) \nonumber \\
& =&\mathbb{1}-i \delta t(1-\hat{h}_z\Big|_{(0,0)})-\left.\delta \mathbf{h}^2\right|_{(0,0)}\left(-\lambda_7^2+P_0^0\right)
\ee
$\delta \mathbf{h}^2\Big|_{(0,0)}$ is calculated as following:
\be
 \mathbf{h}^2&=&\left(\begin{array}{ccc}
\hat{h}_x^2+h_y^2 & \hat{h}_y \hat{h}_z & -\hat{h}_x \hat{h}_z \\
\hat{h}_y \hat{h}_z & \hat{h}_z^2+\hat{h}_x^2 & \hat{h}_x \hat{h}_y \\
-\hat{h}_x \hat{h}_z & \hat{h}_x \hat{h}_y & \hat{h}_y^2+h_z^2
\end{array}\right)  \\
 \left.\delta \mathbf{h}^2\right|_{(0,0)}&\doteq&\left(\begin{array}{ccc}
\delta k_x^2+\delta k_y^2 & \hat{h}_z \delta k_y & -\hat{h}_z \delta k_x \\
\hat{h}_z \delta k_y & \delta k_x^2 & \delta k_x \delta k_y \\
-\hat{h}_z \delta k_x & \delta k_x \delta k_y & \delta k_y
\end{array}\right) \nonumber \\
& \doteq& \hat{h}_z\left(\begin{array}{ccc}
0 & \delta k_y & -\delta k_x \\
\delta k_y & 0 & \\
-\delta k_x & &0
\end{array}\right) 
\ee
Then finally we get
\be
&& \left.U_l\left(\delta t, \delta k_x, \delta k_y\right)\right|_{(\pi, 0,0)} \nonumber\\
 &=&\mathbb{1}+i \delta t \lambda_7\left(\hat{h}_z\Big|_{(0,0)}-1\right)\nonumber \\
&&-\left(\begin{array}{ccc}
	0 & \hat{h}_z \delta k_y & -\hat{h}_z \delta k_x \\
	\hat{h}_z \delta k_y & 0 & \\
	-\hat{h}_z \delta k_x && 0
\end{array}\right)
\left(\begin{array}{ccc}
	1 & & \\
	& -1 &\\
	 && -1
\end{array}\right) \nonumber
 \\
& =&\mathbb{1}+i \delta t \lambda_7\left(\hat{h}_z\Big|_{(0,0)}-1\right)+i \hat{h}_z\Big|_{(0,0)}(\lambda_2 \delta k_y- \lambda_5 \delta k_x )\nonumber\\
& =&\mathbb{1}-i \sgn{(M-2)}( \delta k_x\lambda_5-\delta k_y \lambda_2  )+i \left(\sgn{(M-2)}-1\right)\delta t \lambda_7\nonumber\\
\ee

\section{Phase bands with energy shift}\label{Phasebandenergyshift}
In Sec.(\ref{Mutilfoldfermion}) we showed the three-fold fermion in the phase band with a 2$\pi$ period in time. We can add a chemical potential to shift the overall energy via Eq. (\ref{bandflatmu}). Then the two models become insulators and the period of time in loop unitary is changed back into $\pi$. 
The loop unitary in such case is
 \be
U_l & =&e^{i (\mathbf{h}+P_0) t} e^{-i (\mathbf{h}_0+P_0^0) t}\nonumber \\
& =&\left(\cos t ~\mathbf{h}^2+i \sin t ~(\mathbf{h}+P_0)\right) \cdot\left(\cos t ~\lambda_7^2-i \sin t ~(\lambda_7+P_0^0)\right)\,.\nonumber\\
\ee
However the sacrifice is that the defect is no longer three-fold. We show the plot of phase bands here. 
\begin{figure}[h]
	\centering  %
	\includegraphics[width=0.8\linewidth]{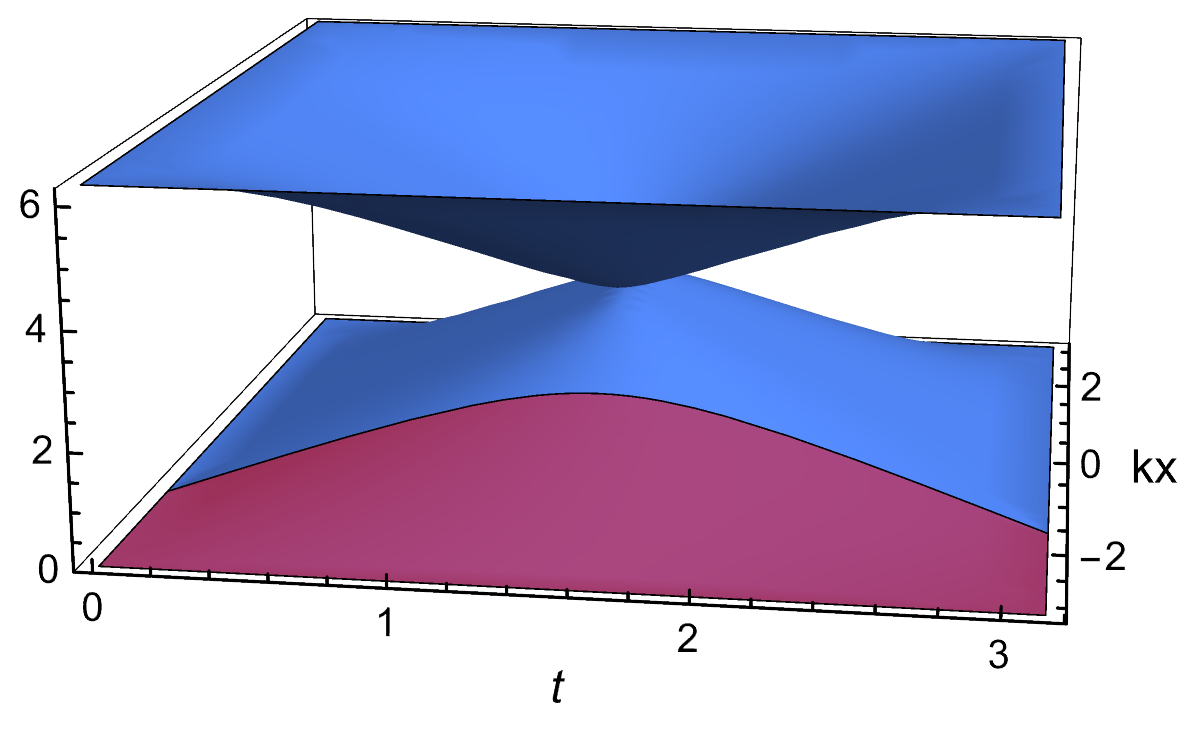} 	
	\caption{Three phase band dispersions at $k_y=0$, we made the second band transparent for the region $-\pi<k_x<-\pi/2$.}  %
	\label{phase3bandpidefect}   %
\end{figure}
{Let us explain the situation here. It is easy to see that at $t=n\pi$ all three phase bands degenerate at $\phi=0$, and therefore to have a flat band which has a constant value, the phase is always zero. For a three-fold fermion to exist, the other two phase bands should meet at $\phi=0$, forming a 0-defect. Next, we show that three-fold 0-defect cannot exist in three-band quenching topological insulators. A three-fold 0-defect from a generic three-band loop unitary should satisfy
\be\label{Ul3band}
	U_l|_{\text{defect}} & =&e^{i \mathbf{h} t} e^{-i \mathbf{h}_0 t}\nonumber \\
	& =&\cos^2 t +\sin^2 t~ \mathbf{h}  \cdot \mathbf{h}_0+\cos t \sin t(\mathbf{h}-\mathbf{h}_0)= \mathbb{1} \,,\nonumber\\
\ee 
which is at least Hermitian.
But $ \mathbf{h}  \cdot \mathbf{h}_0$ is in general non-Hermitian with the exclusion that $\mathbf{h}=\pm \mathbf{h}_0$ at some point in momentum space. The condition that $\mathbf{h}=\mathbf{h}_0$ gives a gapless line in time-momentum space, which is not what we want. The condition that $\mathbf{h}=-\mathbf{h}_0$ and Eq. (\ref{Ul3band}) gives $t=0,\pi/2$. At $t=0$ the phase band is never a cone so no topological structure can appear and at $t=\pi/2$ we have $U_l =-\mathbb{1}$. Therefore, all possibilities are exhausted.
The proof above, however, does not exclude the possibility of multifold fermions without flat phase bands, which can happen for four band cases.}
\bibliography{Loopunitarymultiband}

\begin{thebibliography}{67}%
\makeatletter
\providecommand \@ifxundefined [1]{%
 \@ifx{#1\undefined}
}%
\providecommand \@ifnum [1]{%
 \ifnum #1\expandafter \@firstoftwo
 \else \expandafter \@secondoftwo
 \fi
}%
\providecommand \@ifx [1]{%
 \ifx #1\expandafter \@firstoftwo
 \else \expandafter \@secondoftwo
 \fi
}%
\providecommand \natexlab [1]{#1}%
\providecommand \enquote  [1]{``#1''}%
\providecommand \bibnamefont  [1]{#1}%
\providecommand \bibfnamefont [1]{#1}%
\providecommand \citenamefont [1]{#1}%
\providecommand \href@noop [0]{\@secondoftwo}%
\providecommand \href [0]{\begingroup \@sanitize@url \@href}%
\providecommand \@href[1]{\@@startlink{#1}\@@href}%
\providecommand \@@href[1]{\endgroup#1\@@endlink}%
\providecommand \@sanitize@url [0]{\catcode `\\12\catcode `\$12\catcode
  `\&12\catcode `\#12\catcode `\^12\catcode `\_12\catcode `\%12\relax}%
\providecommand \@@startlink[1]{}%
\providecommand \@@endlink[0]{}%
\providecommand \url  [0]{\begingroup\@sanitize@url \@url }%
\providecommand \@url [1]{\endgroup\@href {#1}{\urlprefix }}%
\providecommand \urlprefix  [0]{URL }%
\providecommand \Eprint [0]{\href }%
\providecommand \doibase [0]{http://dx.doi.org/}%
\providecommand \selectlanguage [0]{\@gobble}%
\providecommand \bibinfo  [0]{\@secondoftwo}%
\providecommand \bibfield  [0]{\@secondoftwo}%
\providecommand \translation [1]{[#1]}%
\providecommand \BibitemOpen [0]{}%
\providecommand \bibitemStop [0]{}%
\providecommand \bibitemNoStop [0]{.\EOS\space}%
\providecommand \EOS [0]{\spacefactor3000\relax}%
\providecommand \BibitemShut  [1]{\csname bibitem#1\endcsname}%
\let\auto@bib@innerbib\@empty
\bibitem [{\citenamefont {Hasan}\ and\ \citenamefont
  {Kane}(2010)}]{hasan2010colloquium}%
  \BibitemOpen
  \bibfield  {author} {\bibinfo {author} {\bibfnamefont {M.~Z.}\ \bibnamefont
  {Hasan}}\ and\ \bibinfo {author} {\bibfnamefont {C.~L.}\ \bibnamefont
  {Kane}},\ }\href {\doibase 10.1103/RevModPhys.82.3045} {\bibfield  {journal}
  {\bibinfo  {journal} {Rev. Mod. Phys.}\ }\textbf {\bibinfo {volume} {82}},\
  \bibinfo {pages} {3045} (\bibinfo {year} {2010})}\BibitemShut {NoStop}%
\bibitem [{\citenamefont {Wen}(2004)}]{Wen:2004ym}%
  \BibitemOpen
  \bibfield  {author} {\bibinfo {author} {\bibfnamefont {X.-G.}\ \bibnamefont
  {Wen}},\ }\href {\doibase 10.1093/acprof:oso/9780199227259.001.0001} {\emph
  {\bibinfo {title} {{Quantum Field Theory of Many-Body Systems: From the
  Origin of Sound to an Origin of Light and Electrons}}}}\ (\bibinfo
  {publisher} {Oxford Univ. Press},\ \bibinfo {year} {2004})\BibitemShut
  {NoStop}%
\bibitem [{\citenamefont {Kitagawa}\ \emph {et~al.}(2010)\citenamefont
  {Kitagawa}, \citenamefont {Berg}, \citenamefont {Rudner},\ and\ \citenamefont
  {Demler}}]{PhysRevB.82.235114}%
  \BibitemOpen
  \bibfield  {author} {\bibinfo {author} {\bibfnamefont {T.}~\bibnamefont
  {Kitagawa}}, \bibinfo {author} {\bibfnamefont {E.}~\bibnamefont {Berg}},
  \bibinfo {author} {\bibfnamefont {M.}~\bibnamefont {Rudner}}, \ and\ \bibinfo
  {author} {\bibfnamefont {E.}~\bibnamefont {Demler}},\ }\href {\doibase
  10.1103/PhysRevB.82.235114} {\bibfield  {journal} {\bibinfo  {journal} {Phys.
  Rev. B}\ }\textbf {\bibinfo {volume} {82}},\ \bibinfo {pages} {235114}
  (\bibinfo {year} {2010})}\BibitemShut {NoStop}%
\bibitem [{\citenamefont {Lindner}\ \emph {et~al.}(2011)\citenamefont
  {Lindner}, \citenamefont {Refael},\ and\ \citenamefont
  {Galitski}}]{Lindner:2011aa}%
  \BibitemOpen
  \bibfield  {author} {\bibinfo {author} {\bibfnamefont {N.~H.}\ \bibnamefont
  {Lindner}}, \bibinfo {author} {\bibfnamefont {G.}~\bibnamefont {Refael}}, \
  and\ \bibinfo {author} {\bibfnamefont {V.}~\bibnamefont {Galitski}},\ }\href
  {\doibase 10.1038/nphys1926} {\bibfield  {journal} {\bibinfo  {journal}
  {Nature Physics}\ }\textbf {\bibinfo {volume} {7}},\ \bibinfo {pages} {490}
  (\bibinfo {year} {2011})}\BibitemShut {NoStop}%
\bibitem [{\citenamefont {Jiang}\ \emph {et~al.}(2011)\citenamefont {Jiang},
  \citenamefont {Kitagawa}, \citenamefont {Alicea}, \citenamefont {Akhmerov},
  \citenamefont {Pekker}, \citenamefont {Refael}, \citenamefont {Cirac},
  \citenamefont {Demler}, \citenamefont {Lukin},\ and\ \citenamefont
  {Zoller}}]{PhysRevLett.106.220402}%
  \BibitemOpen
  \bibfield  {author} {\bibinfo {author} {\bibfnamefont {L.}~\bibnamefont
  {Jiang}}, \bibinfo {author} {\bibfnamefont {T.}~\bibnamefont {Kitagawa}},
  \bibinfo {author} {\bibfnamefont {J.}~\bibnamefont {Alicea}}, \bibinfo
  {author} {\bibfnamefont {A.~R.}\ \bibnamefont {Akhmerov}}, \bibinfo {author}
  {\bibfnamefont {D.}~\bibnamefont {Pekker}}, \bibinfo {author} {\bibfnamefont
  {G.}~\bibnamefont {Refael}}, \bibinfo {author} {\bibfnamefont {J.~I.}\
  \bibnamefont {Cirac}}, \bibinfo {author} {\bibfnamefont {E.}~\bibnamefont
  {Demler}}, \bibinfo {author} {\bibfnamefont {M.~D.}\ \bibnamefont {Lukin}}, \
  and\ \bibinfo {author} {\bibfnamefont {P.}~\bibnamefont {Zoller}},\ }\href
  {\doibase 10.1103/PhysRevLett.106.220402} {\bibfield  {journal} {\bibinfo
  {journal} {Phys. Rev. Lett.}\ }\textbf {\bibinfo {volume} {106}},\ \bibinfo
  {pages} {220402} (\bibinfo {year} {2011})}\BibitemShut {NoStop}%
\bibitem [{\citenamefont {Rudner}\ \emph {et~al.}(2013)\citenamefont {Rudner},
  \citenamefont {Lindner}, \citenamefont {Berg},\ and\ \citenamefont
  {Levin}}]{PhysRevX.3.031005}%
  \BibitemOpen
  \bibfield  {author} {\bibinfo {author} {\bibfnamefont {M.~S.}\ \bibnamefont
  {Rudner}}, \bibinfo {author} {\bibfnamefont {N.~H.}\ \bibnamefont {Lindner}},
  \bibinfo {author} {\bibfnamefont {E.}~\bibnamefont {Berg}}, \ and\ \bibinfo
  {author} {\bibfnamefont {M.}~\bibnamefont {Levin}},\ }\href {\doibase
  10.1103/PhysRevX.3.031005} {\bibfield  {journal} {\bibinfo  {journal} {Phys.
  Rev. X}\ }\textbf {\bibinfo {volume} {3}},\ \bibinfo {pages} {031005}
  (\bibinfo {year} {2013})}\BibitemShut {NoStop}%
\bibitem [{\citenamefont {G\'omez-Le\'on}\ and\ \citenamefont
  {Platero}(2013)}]{PhysRevLett.110.200403}%
  \BibitemOpen
  \bibfield  {author} {\bibinfo {author} {\bibfnamefont {A.}~\bibnamefont
  {G\'omez-Le\'on}}\ and\ \bibinfo {author} {\bibfnamefont {G.}~\bibnamefont
  {Platero}},\ }\href {\doibase 10.1103/PhysRevLett.110.200403} {\bibfield
  {journal} {\bibinfo  {journal} {Phys. Rev. Lett.}\ }\textbf {\bibinfo
  {volume} {110}},\ \bibinfo {pages} {200403} (\bibinfo {year}
  {2013})}\BibitemShut {NoStop}%
\bibitem [{\citenamefont {D'Alessio}\ and\ \citenamefont
  {Rigol}(2015)}]{DAlessio:2015aa}%
  \BibitemOpen
  \bibfield  {author} {\bibinfo {author} {\bibfnamefont {L.}~\bibnamefont
  {D'Alessio}}\ and\ \bibinfo {author} {\bibfnamefont {M.}~\bibnamefont
  {Rigol}},\ }\href {\doibase 10.1038/ncomms9336} {\bibfield  {journal}
  {\bibinfo  {journal} {Nature Communications}\ }\textbf {\bibinfo {volume}
  {6}},\ \bibinfo {pages} {8336} (\bibinfo {year} {2015})}\BibitemShut
  {NoStop}%
\bibitem [{\citenamefont {Potter}\ \emph {et~al.}(2016)\citenamefont {Potter},
  \citenamefont {Morimoto},\ and\ \citenamefont
  {Vishwanath}}]{PhysRevX.6.041001}%
  \BibitemOpen
  \bibfield  {author} {\bibinfo {author} {\bibfnamefont {A.~C.}\ \bibnamefont
  {Potter}}, \bibinfo {author} {\bibfnamefont {T.}~\bibnamefont {Morimoto}}, \
  and\ \bibinfo {author} {\bibfnamefont {A.}~\bibnamefont {Vishwanath}},\
  }\href {\doibase 10.1103/PhysRevX.6.041001} {\bibfield  {journal} {\bibinfo
  {journal} {Phys. Rev. X}\ }\textbf {\bibinfo {volume} {6}},\ \bibinfo {pages}
  {041001} (\bibinfo {year} {2016})}\BibitemShut {NoStop}%
\bibitem [{\citenamefont {Higashikawa}\ \emph {et~al.}(2019)\citenamefont
  {Higashikawa}, \citenamefont {Nakagawa},\ and\ \citenamefont
  {Ueda}}]{PhysRevLett.123.066403}%
  \BibitemOpen
  \bibfield  {author} {\bibinfo {author} {\bibfnamefont {S.}~\bibnamefont
  {Higashikawa}}, \bibinfo {author} {\bibfnamefont {M.}~\bibnamefont
  {Nakagawa}}, \ and\ \bibinfo {author} {\bibfnamefont {M.}~\bibnamefont
  {Ueda}},\ }\href {\doibase 10.1103/PhysRevLett.123.066403} {\bibfield
  {journal} {\bibinfo  {journal} {Phys. Rev. Lett.}\ }\textbf {\bibinfo
  {volume} {123}},\ \bibinfo {pages} {066403} (\bibinfo {year}
  {2019})}\BibitemShut {NoStop}%
\bibitem [{\citenamefont {Hu}\ \emph {et~al.}(2020{\natexlab{a}})\citenamefont
  {Hu}, \citenamefont {Huang}, \citenamefont {Zhao},\ and\ \citenamefont
  {Liu}}]{PhysRevLett.124.057001}%
  \BibitemOpen
  \bibfield  {author} {\bibinfo {author} {\bibfnamefont {H.}~\bibnamefont
  {Hu}}, \bibinfo {author} {\bibfnamefont {B.}~\bibnamefont {Huang}}, \bibinfo
  {author} {\bibfnamefont {E.}~\bibnamefont {Zhao}}, \ and\ \bibinfo {author}
  {\bibfnamefont {W.~V.}\ \bibnamefont {Liu}},\ }\href {\doibase
  10.1103/PhysRevLett.124.057001} {\bibfield  {journal} {\bibinfo  {journal}
  {Phys. Rev. Lett.}\ }\textbf {\bibinfo {volume} {124}},\ \bibinfo {pages}
  {057001} (\bibinfo {year} {2020}{\natexlab{a}})}\BibitemShut {NoStop}%
\bibitem [{\citenamefont {Roy}\ and\ \citenamefont
  {Harper}(2017)}]{PhysRevB.96.155118}%
  \BibitemOpen
  \bibfield  {author} {\bibinfo {author} {\bibfnamefont {R.}~\bibnamefont
  {Roy}}\ and\ \bibinfo {author} {\bibfnamefont {F.}~\bibnamefont {Harper}},\
  }\href {\doibase 10.1103/PhysRevB.96.155118} {\bibfield  {journal} {\bibinfo
  {journal} {Phys. Rev. B}\ }\textbf {\bibinfo {volume} {96}},\ \bibinfo
  {pages} {155118} (\bibinfo {year} {2017})}\BibitemShut {NoStop}%
\bibitem [{\citenamefont {Yao}\ \emph {et~al.}(2017)\citenamefont {Yao},
  \citenamefont {Yan},\ and\ \citenamefont {Wang}}]{PhysRevB.96.195303}%
  \BibitemOpen
  \bibfield  {author} {\bibinfo {author} {\bibfnamefont {S.}~\bibnamefont
  {Yao}}, \bibinfo {author} {\bibfnamefont {Z.}~\bibnamefont {Yan}}, \ and\
  \bibinfo {author} {\bibfnamefont {Z.}~\bibnamefont {Wang}},\ }\href {\doibase
  10.1103/PhysRevB.96.195303} {\bibfield  {journal} {\bibinfo  {journal} {Phys.
  Rev. B}\ }\textbf {\bibinfo {volume} {96}},\ \bibinfo {pages} {195303}
  (\bibinfo {year} {2017})}\BibitemShut {NoStop}%
\bibitem [{\citenamefont {Yang}\ \emph {et~al.}(2019)\citenamefont {Yang},
  \citenamefont {Zhou}, \citenamefont {Ma}, \citenamefont {Kong}, \citenamefont
  {Wang}, \citenamefont {Qin}, \citenamefont {Rong}, \citenamefont {Wang},
  \citenamefont {Shi}, \citenamefont {Gong},\ and\ \citenamefont
  {Du}}]{PhysRevB.100.085308}%
  \BibitemOpen
  \bibfield  {author} {\bibinfo {author} {\bibfnamefont {K.}~\bibnamefont
  {Yang}}, \bibinfo {author} {\bibfnamefont {L.}~\bibnamefont {Zhou}}, \bibinfo
  {author} {\bibfnamefont {W.}~\bibnamefont {Ma}}, \bibinfo {author}
  {\bibfnamefont {X.}~\bibnamefont {Kong}}, \bibinfo {author} {\bibfnamefont
  {P.}~\bibnamefont {Wang}}, \bibinfo {author} {\bibfnamefont {X.}~\bibnamefont
  {Qin}}, \bibinfo {author} {\bibfnamefont {X.}~\bibnamefont {Rong}}, \bibinfo
  {author} {\bibfnamefont {Y.}~\bibnamefont {Wang}}, \bibinfo {author}
  {\bibfnamefont {F.}~\bibnamefont {Shi}}, \bibinfo {author} {\bibfnamefont
  {J.}~\bibnamefont {Gong}}, \ and\ \bibinfo {author} {\bibfnamefont
  {J.}~\bibnamefont {Du}},\ }\href {\doibase 10.1103/PhysRevB.100.085308}
  {\bibfield  {journal} {\bibinfo  {journal} {Phys. Rev. B}\ }\textbf {\bibinfo
  {volume} {100}},\ \bibinfo {pages} {085308} (\bibinfo {year}
  {2019})}\BibitemShut {NoStop}%
\bibitem [{\citenamefont {Li}\ and\ \citenamefont {Hu}(2023)}]{Li:2023aa}%
  \BibitemOpen
  \bibfield  {author} {\bibinfo {author} {\bibfnamefont {T.}~\bibnamefont
  {Li}}\ and\ \bibinfo {author} {\bibfnamefont {H.}~\bibnamefont {Hu}},\ }\href
  {\doibase 10.1038/s41467-023-42139-z} {\bibfield  {journal} {\bibinfo
  {journal} {Nature Communications}\ }\textbf {\bibinfo {volume} {14}},\
  \bibinfo {pages} {6418} (\bibinfo {year} {2023})}\BibitemShut {NoStop}%
\bibitem [{\citenamefont {Jangjan}\ and\ \citenamefont
  {Hosseini}(2020)}]{Jangjan:2020aa}%
  \BibitemOpen
  \bibfield  {author} {\bibinfo {author} {\bibfnamefont {M.}~\bibnamefont
  {Jangjan}}\ and\ \bibinfo {author} {\bibfnamefont {M.~V.}\ \bibnamefont
  {Hosseini}},\ }\href {\doibase 10.1038/s41598-020-71196-3} {\bibfield
  {journal} {\bibinfo  {journal} {Scientific Reports}\ }\textbf {\bibinfo
  {volume} {10}},\ \bibinfo {pages} {14256} (\bibinfo {year}
  {2020})}\BibitemShut {NoStop}%
\bibitem [{\citenamefont {Jangjan}\ \emph {et~al.}(2022)\citenamefont
  {Jangjan}, \citenamefont {Foa~Torres},\ and\ \citenamefont
  {Hosseini}}]{PhysRevB.106.224306}%
  \BibitemOpen
  \bibfield  {author} {\bibinfo {author} {\bibfnamefont {M.}~\bibnamefont
  {Jangjan}}, \bibinfo {author} {\bibfnamefont {L.~E.~F.}\ \bibnamefont
  {Foa~Torres}}, \ and\ \bibinfo {author} {\bibfnamefont {M.~V.}\ \bibnamefont
  {Hosseini}},\ }\href {\doibase 10.1103/PhysRevB.106.224306} {\bibfield
  {journal} {\bibinfo  {journal} {Phys. Rev. B}\ }\textbf {\bibinfo {volume}
  {106}},\ \bibinfo {pages} {224306} (\bibinfo {year} {2022})}\BibitemShut
  {NoStop}%
\bibitem [{\citenamefont {Fl{\"a}schner}\ \emph {et~al.}(2016)\citenamefont
  {Fl{\"a}schner}, \citenamefont {Rem}, \citenamefont {Tarnowski},
  \citenamefont {Vogel}, \citenamefont {L{\"u}hmann}, \citenamefont
  {Sengstock},\ and\ \citenamefont {Weitenberg}}]{doi:10.1126/science.aad4568}%
  \BibitemOpen
  \bibfield  {author} {\bibinfo {author} {\bibfnamefont {N.}~\bibnamefont
  {Fl{\"a}schner}}, \bibinfo {author} {\bibfnamefont {B.~S.}\ \bibnamefont
  {Rem}}, \bibinfo {author} {\bibfnamefont {M.}~\bibnamefont {Tarnowski}},
  \bibinfo {author} {\bibfnamefont {D.}~\bibnamefont {Vogel}}, \bibinfo
  {author} {\bibfnamefont {D.-S.}\ \bibnamefont {L{\"u}hmann}}, \bibinfo
  {author} {\bibfnamefont {K.}~\bibnamefont {Sengstock}}, \ and\ \bibinfo
  {author} {\bibfnamefont {C.}~\bibnamefont {Weitenberg}},\ }\href {\doibase
  10.1126/science.aad4568} {\bibfield  {journal} {\bibinfo  {journal}
  {Science}\ }\textbf {\bibinfo {volume} {352}},\ \bibinfo {pages} {1091}
  (\bibinfo {year} {2016})},\ \Eprint
  {http://arxiv.org/abs/https://www.science.org/doi/pdf/10.1126/science.aad4568}
  {https://www.science.org/doi/pdf/10.1126/science.aad4568} \BibitemShut
  {NoStop}%
\bibitem [{\citenamefont {Alba}\ \emph {et~al.}(2011)\citenamefont {Alba},
  \citenamefont {Fernandez-Gonzalvo}, \citenamefont {Mur-Petit}, \citenamefont
  {Pachos},\ and\ \citenamefont {Garcia-Ripoll}}]{PhysRevLett.107.235301}%
  \BibitemOpen
  \bibfield  {author} {\bibinfo {author} {\bibfnamefont {E.}~\bibnamefont
  {Alba}}, \bibinfo {author} {\bibfnamefont {X.}~\bibnamefont
  {Fernandez-Gonzalvo}}, \bibinfo {author} {\bibfnamefont {J.}~\bibnamefont
  {Mur-Petit}}, \bibinfo {author} {\bibfnamefont {J.~K.}\ \bibnamefont
  {Pachos}}, \ and\ \bibinfo {author} {\bibfnamefont {J.~J.}\ \bibnamefont
  {Garcia-Ripoll}},\ }\href {\doibase 10.1103/PhysRevLett.107.235301}
  {\bibfield  {journal} {\bibinfo  {journal} {Phys. Rev. Lett.}\ }\textbf
  {\bibinfo {volume} {107}},\ \bibinfo {pages} {235301} (\bibinfo {year}
  {2011})}\BibitemShut {NoStop}%
\bibitem [{\citenamefont {Hauke}\ \emph {et~al.}(2014)\citenamefont {Hauke},
  \citenamefont {Lewenstein},\ and\ \citenamefont
  {Eckardt}}]{PhysRevLett.113.045303}%
  \BibitemOpen
  \bibfield  {author} {\bibinfo {author} {\bibfnamefont {P.}~\bibnamefont
  {Hauke}}, \bibinfo {author} {\bibfnamefont {M.}~\bibnamefont {Lewenstein}}, \
  and\ \bibinfo {author} {\bibfnamefont {A.}~\bibnamefont {Eckardt}},\ }\href
  {\doibase 10.1103/PhysRevLett.113.045303} {\bibfield  {journal} {\bibinfo
  {journal} {Phys. Rev. Lett.}\ }\textbf {\bibinfo {volume} {113}},\ \bibinfo
  {pages} {045303} (\bibinfo {year} {2014})}\BibitemShut {NoStop}%
\bibitem [{\citenamefont {Caio}\ \emph {et~al.}(2015)\citenamefont {Caio},
  \citenamefont {Cooper},\ and\ \citenamefont {Bhaseen}}]{CaioMD2015}%
  \BibitemOpen
  \bibfield  {author} {\bibinfo {author} {\bibfnamefont {M.~D.}\ \bibnamefont
  {Caio}}, \bibinfo {author} {\bibfnamefont {N.~R.}\ \bibnamefont {Cooper}}, \
  and\ \bibinfo {author} {\bibfnamefont {M.~J.}\ \bibnamefont {Bhaseen}},\
  }\href {\doibase 10.1103/PhysRevLett.115.236403} {\bibfield  {journal}
  {\bibinfo  {journal} {Phys. Rev. Lett.}\ }\textbf {\bibinfo {volume} {115}},\
  \bibinfo {pages} {236403} (\bibinfo {year} {2015})}\BibitemShut {NoStop}%
\bibitem [{\citenamefont {Caio}\ \emph {et~al.}(2016)\citenamefont {Caio},
  \citenamefont {Cooper},\ and\ \citenamefont {Bhaseen}}]{PhysRevB.94.155104}%
  \BibitemOpen
  \bibfield  {author} {\bibinfo {author} {\bibfnamefont {M.~D.}\ \bibnamefont
  {Caio}}, \bibinfo {author} {\bibfnamefont {N.~R.}\ \bibnamefont {Cooper}}, \
  and\ \bibinfo {author} {\bibfnamefont {M.~J.}\ \bibnamefont {Bhaseen}},\
  }\href {\doibase 10.1103/PhysRevB.94.155104} {\bibfield  {journal} {\bibinfo
  {journal} {Phys. Rev. B}\ }\textbf {\bibinfo {volume} {94}},\ \bibinfo
  {pages} {155104} (\bibinfo {year} {2016})}\BibitemShut {NoStop}%
\bibitem [{\citenamefont {Hu}\ \emph {et~al.}(2016)\citenamefont {Hu},
  \citenamefont {Zoller},\ and\ \citenamefont {Budich}}]{Ying2016}%
  \BibitemOpen
  \bibfield  {author} {\bibinfo {author} {\bibfnamefont {Y.}~\bibnamefont
  {Hu}}, \bibinfo {author} {\bibfnamefont {P.}~\bibnamefont {Zoller}}, \ and\
  \bibinfo {author} {\bibfnamefont {J.~C.}\ \bibnamefont {Budich}},\ }\href
  {\doibase 10.1103/PhysRevLett.117.126803} {\bibfield  {journal} {\bibinfo
  {journal} {Phys. Rev. Lett.}\ }\textbf {\bibinfo {volume} {117}},\ \bibinfo
  {pages} {126803} (\bibinfo {year} {2016})}\BibitemShut {NoStop}%
\bibitem [{\citenamefont {Tarnowski}\ \emph {et~al.}(2019)\citenamefont
  {Tarnowski}, \citenamefont {{\"U}nal}, \citenamefont {Fl{\"a}schner},
  \citenamefont {Rem}, \citenamefont {Eckardt}, \citenamefont {Sengstock},\
  and\ \citenamefont {Weitenberg}}]{Tarnowski:2019aa}%
  \BibitemOpen
  \bibfield  {author} {\bibinfo {author} {\bibfnamefont {M.}~\bibnamefont
  {Tarnowski}}, \bibinfo {author} {\bibfnamefont {F.~N.}\ \bibnamefont
  {{\"U}nal}}, \bibinfo {author} {\bibfnamefont {N.}~\bibnamefont
  {Fl{\"a}schner}}, \bibinfo {author} {\bibfnamefont {B.~S.}\ \bibnamefont
  {Rem}}, \bibinfo {author} {\bibfnamefont {A.}~\bibnamefont {Eckardt}},
  \bibinfo {author} {\bibfnamefont {K.}~\bibnamefont {Sengstock}}, \ and\
  \bibinfo {author} {\bibfnamefont {C.}~\bibnamefont {Weitenberg}},\ }\href
  {\doibase 10.1038/s41467-019-09668-y} {\bibfield  {journal} {\bibinfo
  {journal} {Nature Communications}\ }\textbf {\bibinfo {volume} {10}},\
  \bibinfo {pages} {1728} (\bibinfo {year} {2019})}\BibitemShut {NoStop}%
\bibitem [{\citenamefont {\"Unal}\ \emph {et~al.}(2020)\citenamefont {\"Unal},
  \citenamefont {Bouhon},\ and\ \citenamefont
  {Slager}}]{PhysRevLett.125.053601}%
  \BibitemOpen
  \bibfield  {author} {\bibinfo {author} {\bibfnamefont {F.~N.}\ \bibnamefont
  {\"Unal}}, \bibinfo {author} {\bibfnamefont {A.}~\bibnamefont {Bouhon}}, \
  and\ \bibinfo {author} {\bibfnamefont {R.-J.}\ \bibnamefont {Slager}},\
  }\href {\doibase 10.1103/PhysRevLett.125.053601} {\bibfield  {journal}
  {\bibinfo  {journal} {Phys. Rev. Lett.}\ }\textbf {\bibinfo {volume} {125}},\
  \bibinfo {pages} {053601} (\bibinfo {year} {2020})}\BibitemShut {NoStop}%
\bibitem [{\citenamefont {Ul\ifmmode~\check{c}\else \v{c}\fi{}akar}\ \emph
  {et~al.}(2018)\citenamefont {Ul\ifmmode~\check{c}\else \v{c}\fi{}akar},
  \citenamefont {Mravlje}, \citenamefont {Ram\ifmmode~\check{s}\else
  \v{s}\fi{}ak},\ and\ \citenamefont {Rejec}}]{PhysRevB.97.195127}%
  \BibitemOpen
  \bibfield  {author} {\bibinfo {author} {\bibfnamefont {L.}~\bibnamefont
  {Ul\ifmmode~\check{c}\else \v{c}\fi{}akar}}, \bibinfo {author} {\bibfnamefont
  {J.}~\bibnamefont {Mravlje}}, \bibinfo {author} {\bibfnamefont
  {A.}~\bibnamefont {Ram\ifmmode~\check{s}\else \v{s}\fi{}ak}}, \ and\ \bibinfo
  {author} {\bibfnamefont {T.~c.~v.}\ \bibnamefont {Rejec}},\ }\href {\doibase
  10.1103/PhysRevB.97.195127} {\bibfield  {journal} {\bibinfo  {journal} {Phys.
  Rev. B}\ }\textbf {\bibinfo {volume} {97}},\ \bibinfo {pages} {195127}
  (\bibinfo {year} {2018})}\BibitemShut {NoStop}%
\bibitem [{\citenamefont {Zhang}\ \emph {et~al.}(2018)\citenamefont {Zhang},
  \citenamefont {Zhang}, \citenamefont {Niu},\ and\ \citenamefont
  {Liu}}]{ZHANG20181385}%
  \BibitemOpen
  \bibfield  {author} {\bibinfo {author} {\bibfnamefont {L.}~\bibnamefont
  {Zhang}}, \bibinfo {author} {\bibfnamefont {L.}~\bibnamefont {Zhang}},
  \bibinfo {author} {\bibfnamefont {S.}~\bibnamefont {Niu}}, \ and\ \bibinfo
  {author} {\bibfnamefont {X.-J.}\ \bibnamefont {Liu}},\ }\href {\doibase
  https://doi.org/10.1016/j.scib.2018.09.018} {\bibfield  {journal} {\bibinfo
  {journal} {Sci. Bull.}\ }\textbf {\bibinfo {volume} {63}},\ \bibinfo {pages}
  {1385} (\bibinfo {year} {2018})}\BibitemShut {NoStop}%
\bibitem [{\citenamefont {Zhang}\ \emph {et~al.}(2019)\citenamefont {Zhang},
  \citenamefont {Zhang},\ and\ \citenamefont {Liu}}]{ZhangLong2019}%
  \BibitemOpen
  \bibfield  {author} {\bibinfo {author} {\bibfnamefont {L.}~\bibnamefont
  {Zhang}}, \bibinfo {author} {\bibfnamefont {L.}~\bibnamefont {Zhang}}, \ and\
  \bibinfo {author} {\bibfnamefont {X.-J.}\ \bibnamefont {Liu}},\ }\href
  {\doibase 10.1103/PhysRevA.99.053606} {\bibfield  {journal} {\bibinfo
  {journal} {Phys. Rev. A.}\ }\textbf {\bibinfo {volume} {99}},\ \bibinfo
  {pages} {053606} (\bibinfo {year} {2019})}\BibitemShut {NoStop}%
\bibitem [{\citenamefont {Wang}\ \emph {et~al.}(2019)\citenamefont {Wang},
  \citenamefont {Ji}, \citenamefont {Chai}, \citenamefont {Guo}, \citenamefont
  {Wang}, \citenamefont {Ye}, \citenamefont {Yu}, \citenamefont {Zhang},
  \citenamefont {Qin}, \citenamefont {Wang}, \citenamefont {Shi}, \citenamefont
  {Rong}, \citenamefont {Lu}, \citenamefont {Liu},\ and\ \citenamefont
  {Du}}]{YaWangXJLiu2019}%
  \BibitemOpen
  \bibfield  {author} {\bibinfo {author} {\bibfnamefont {Y.}~\bibnamefont
  {Wang}}, \bibinfo {author} {\bibfnamefont {W.}~\bibnamefont {Ji}}, \bibinfo
  {author} {\bibfnamefont {Z.}~\bibnamefont {Chai}}, \bibinfo {author}
  {\bibfnamefont {Y.}~\bibnamefont {Guo}}, \bibinfo {author} {\bibfnamefont
  {M.}~\bibnamefont {Wang}}, \bibinfo {author} {\bibfnamefont {X.}~\bibnamefont
  {Ye}}, \bibinfo {author} {\bibfnamefont {P.}~\bibnamefont {Yu}}, \bibinfo
  {author} {\bibfnamefont {L.}~\bibnamefont {Zhang}}, \bibinfo {author}
  {\bibfnamefont {X.}~\bibnamefont {Qin}}, \bibinfo {author} {\bibfnamefont
  {P.}~\bibnamefont {Wang}}, \bibinfo {author} {\bibfnamefont {F.}~\bibnamefont
  {Shi}}, \bibinfo {author} {\bibfnamefont {X.}~\bibnamefont {Rong}}, \bibinfo
  {author} {\bibfnamefont {D.}~\bibnamefont {Lu}}, \bibinfo {author}
  {\bibfnamefont {X.-J.}\ \bibnamefont {Liu}}, \ and\ \bibinfo {author}
  {\bibfnamefont {J.}~\bibnamefont {Du}},\ }\href {\doibase
  10.1103/PhysRevA.100.052328} {\bibfield  {journal} {\bibinfo  {journal}
  {Phys. Rev. A.}\ }\textbf {\bibinfo {volume} {100}},\ \bibinfo {pages}
  {052328} (\bibinfo {year} {2019})}\BibitemShut {NoStop}%
\bibitem [{\citenamefont {Yu}\ \emph {et~al.}(2021)\citenamefont {Yu},
  \citenamefont {Ji}, \citenamefont {Zhang}, \citenamefont {Wang},
  \citenamefont {Wu},\ and\ \citenamefont {Liu}}]{YuXiangLong2021}%
  \BibitemOpen
  \bibfield  {author} {\bibinfo {author} {\bibfnamefont {X.-L.}\ \bibnamefont
  {Yu}}, \bibinfo {author} {\bibfnamefont {W.}~\bibnamefont {Ji}}, \bibinfo
  {author} {\bibfnamefont {L.}~\bibnamefont {Zhang}}, \bibinfo {author}
  {\bibfnamefont {Y.}~\bibnamefont {Wang}}, \bibinfo {author} {\bibfnamefont
  {J.}~\bibnamefont {Wu}}, \ and\ \bibinfo {author} {\bibfnamefont {X.-J.}\
  \bibnamefont {Liu}},\ }\href {\doibase 10.1103/PRXQuantum.2.020320}
  {\bibfield  {journal} {\bibinfo  {journal} {PRX Quantum}\ }\textbf {\bibinfo
  {volume} {2}},\ \bibinfo {pages} {020320} (\bibinfo {year}
  {2021})}\BibitemShut {NoStop}%
\bibitem [{\citenamefont {Zhang}\ \emph {et~al.}(2020)\citenamefont {Zhang},
  \citenamefont {Zhang},\ and\ \citenamefont {Liu}}]{ZhangLong2020}%
  \BibitemOpen
  \bibfield  {author} {\bibinfo {author} {\bibfnamefont {L.}~\bibnamefont
  {Zhang}}, \bibinfo {author} {\bibfnamefont {L.}~\bibnamefont {Zhang}}, \ and\
  \bibinfo {author} {\bibfnamefont {X.-J.}\ \bibnamefont {Liu}},\ }\href
  {\doibase 10.1103/PhysRevLett.125.183001} {\bibfield  {journal} {\bibinfo
  {journal} {Phys. Rev. Lett.}\ }\textbf {\bibinfo {volume} {125}},\ \bibinfo
  {pages} {183001} (\bibinfo {year} {2020})}\BibitemShut {NoStop}%
\bibitem [{\citenamefont {Sun}\ \emph {et~al.}(2018)\citenamefont {Sun},
  \citenamefont {Yi}, \citenamefont {Wang}, \citenamefont {Zhang},
  \citenamefont {Sanders}, \citenamefont {Xu}, \citenamefont {Wang},
  \citenamefont {Schmiedmayer}, \citenamefont {Deng}, \citenamefont {Liu},
  \citenamefont {Chen},\ and\ \citenamefont {Pan}}]{ShuaiChen2018}%
  \BibitemOpen
  \bibfield  {author} {\bibinfo {author} {\bibfnamefont {W.}~\bibnamefont
  {Sun}}, \bibinfo {author} {\bibfnamefont {C.-R.}\ \bibnamefont {Yi}},
  \bibinfo {author} {\bibfnamefont {B.-Z.}\ \bibnamefont {Wang}}, \bibinfo
  {author} {\bibfnamefont {W.-W.}\ \bibnamefont {Zhang}}, \bibinfo {author}
  {\bibfnamefont {B.~C.}\ \bibnamefont {Sanders}}, \bibinfo {author}
  {\bibfnamefont {X.-T.}\ \bibnamefont {Xu}}, \bibinfo {author} {\bibfnamefont
  {Z.-Y.}\ \bibnamefont {Wang}}, \bibinfo {author} {\bibfnamefont
  {J.}~\bibnamefont {Schmiedmayer}}, \bibinfo {author} {\bibfnamefont
  {Y.}~\bibnamefont {Deng}}, \bibinfo {author} {\bibfnamefont {X.-J.}\
  \bibnamefont {Liu}}, \bibinfo {author} {\bibfnamefont {S.}~\bibnamefont
  {Chen}}, \ and\ \bibinfo {author} {\bibfnamefont {J.-W.}\ \bibnamefont
  {Pan}},\ }\href {\doibase 10.1103/PhysRevLett.121.250403} {\bibfield
  {journal} {\bibinfo  {journal} {Phys. Rev. Lett.}\ }\textbf {\bibinfo
  {volume} {121}},\ \bibinfo {pages} {250403} (\bibinfo {year}
  {2018})}\BibitemShut {NoStop}%
\bibitem [{\citenamefont {Li}\ \emph {et~al.}(2021)\citenamefont {Li},
  \citenamefont {Zhu},\ and\ \citenamefont {Gong}}]{LI20211502}%
  \BibitemOpen
  \bibfield  {author} {\bibinfo {author} {\bibfnamefont {L.}~\bibnamefont
  {Li}}, \bibinfo {author} {\bibfnamefont {W.}~\bibnamefont {Zhu}}, \ and\
  \bibinfo {author} {\bibfnamefont {J.}~\bibnamefont {Gong}},\ }\href {\doibase
  https://doi.org/10.1016/j.scib.2021.04.006} {\bibfield  {journal} {\bibinfo
  {journal} {Science Bulletin}\ }\textbf {\bibinfo {volume} {66}},\ \bibinfo
  {pages} {1502} (\bibinfo {year} {2021})}\BibitemShut {NoStop}%
\bibitem [{\citenamefont {Wu}\ \emph {et~al.}(2023)\citenamefont {Wu},
  \citenamefont {Fang},\ and\ \citenamefont {Li}}]{PhysRevA.107.052209}%
  \BibitemOpen
  \bibfield  {author} {\bibinfo {author} {\bibfnamefont {X.}~\bibnamefont
  {Wu}}, \bibinfo {author} {\bibfnamefont {P.}~\bibnamefont {Fang}}, \ and\
  \bibinfo {author} {\bibfnamefont {F.}~\bibnamefont {Li}},\ }\href {\doibase
  10.1103/PhysRevA.107.052209} {\bibfield  {journal} {\bibinfo  {journal}
  {Phys. Rev. A}\ }\textbf {\bibinfo {volume} {107}},\ \bibinfo {pages}
  {052209} (\bibinfo {year} {2023})}\BibitemShut {NoStop}%
\bibitem [{\citenamefont {Zhu}\ \emph {et~al.}(2020)\citenamefont {Zhu},
  \citenamefont {Ke}, \citenamefont {Zhong},\ and\ \citenamefont
  {Lee}}]{PhysRevResearch.2.023043}%
  \BibitemOpen
  \bibfield  {author} {\bibinfo {author} {\bibfnamefont {B.}~\bibnamefont
  {Zhu}}, \bibinfo {author} {\bibfnamefont {Y.}~\bibnamefont {Ke}}, \bibinfo
  {author} {\bibfnamefont {H.}~\bibnamefont {Zhong}}, \ and\ \bibinfo {author}
  {\bibfnamefont {C.}~\bibnamefont {Lee}},\ }\href {\doibase
  10.1103/PhysRevResearch.2.023043} {\bibfield  {journal} {\bibinfo  {journal}
  {Phys. Rev. Res.}\ }\textbf {\bibinfo {volume} {2}},\ \bibinfo {pages}
  {023043} (\bibinfo {year} {2020})}\BibitemShut {NoStop}%
\bibitem [{\citenamefont {Ye}\ and\ \citenamefont
  {Li}(2020)}]{PhysRevA.102.042209}%
  \BibitemOpen
  \bibfield  {author} {\bibinfo {author} {\bibfnamefont {J.}~\bibnamefont
  {Ye}}\ and\ \bibinfo {author} {\bibfnamefont {F.}~\bibnamefont {Li}},\ }\href
  {\doibase 10.1103/PhysRevA.102.042209} {\bibfield  {journal} {\bibinfo
  {journal} {Phys. Rev. A}\ }\textbf {\bibinfo {volume} {102}},\ \bibinfo
  {pages} {042209} (\bibinfo {year} {2020})}\BibitemShut {NoStop}%
\bibitem [{\citenamefont {Fang}\ \emph {et~al.}(2022)\citenamefont {Fang},
  \citenamefont {Wang},\ and\ \citenamefont {Li}}]{PhysRevA.106.022219}%
  \BibitemOpen
  \bibfield  {author} {\bibinfo {author} {\bibfnamefont {P.}~\bibnamefont
  {Fang}}, \bibinfo {author} {\bibfnamefont {Y.-X.}\ \bibnamefont {Wang}}, \
  and\ \bibinfo {author} {\bibfnamefont {F.}~\bibnamefont {Li}},\ }\href
  {\doibase 10.1103/PhysRevA.106.022219} {\bibfield  {journal} {\bibinfo
  {journal} {Phys. Rev. A}\ }\textbf {\bibinfo {volume} {106}},\ \bibinfo
  {pages} {022219} (\bibinfo {year} {2022})}\BibitemShut {NoStop}%
\bibitem [{\citenamefont {Yang}\ \emph {et~al.}(2018)\citenamefont {Yang},
  \citenamefont {Li},\ and\ \citenamefont {Chen}}]{PhysRevB.97.060304}%
  \BibitemOpen
  \bibfield  {author} {\bibinfo {author} {\bibfnamefont {C.}~\bibnamefont
  {Yang}}, \bibinfo {author} {\bibfnamefont {L.}~\bibnamefont {Li}}, \ and\
  \bibinfo {author} {\bibfnamefont {S.}~\bibnamefont {Chen}},\ }\href {\doibase
  10.1103/PhysRevB.97.060304} {\bibfield  {journal} {\bibinfo  {journal} {Phys.
  Rev. B}\ }\textbf {\bibinfo {volume} {97}},\ \bibinfo {pages} {060304}
  (\bibinfo {year} {2018})}\BibitemShut {NoStop}%
\bibitem [{\citenamefont {Wang}\ \emph {et~al.}(2017)\citenamefont {Wang},
  \citenamefont {Zhang}, \citenamefont {Chen}, \citenamefont {Yu},\ and\
  \citenamefont {Zhai}}]{WangCe2017}%
  \BibitemOpen
  \bibfield  {author} {\bibinfo {author} {\bibfnamefont {C.}~\bibnamefont
  {Wang}}, \bibinfo {author} {\bibfnamefont {P.}~\bibnamefont {Zhang}},
  \bibinfo {author} {\bibfnamefont {X.}~\bibnamefont {Chen}}, \bibinfo {author}
  {\bibfnamefont {J.}~\bibnamefont {Yu}}, \ and\ \bibinfo {author}
  {\bibfnamefont {H.}~\bibnamefont {Zhai}},\ }\href {\doibase
  10.1103/PhysRevLett.118.185701} {\bibfield  {journal} {\bibinfo  {journal}
  {Phys. Rev. Lett.}\ }\textbf {\bibinfo {volume} {118}},\ \bibinfo {pages}
  {185701} (\bibinfo {year} {2017})}\BibitemShut {NoStop}%
\bibitem [{\citenamefont {Chen}\ \emph {et~al.}(2020)\citenamefont {Chen},
  \citenamefont {Wang},\ and\ \citenamefont {Yu}}]{ChenXin2020}%
  \BibitemOpen
  \bibfield  {author} {\bibinfo {author} {\bibfnamefont {X.}~\bibnamefont
  {Chen}}, \bibinfo {author} {\bibfnamefont {C.}~\bibnamefont {Wang}}, \ and\
  \bibinfo {author} {\bibfnamefont {J.}~\bibnamefont {Yu}},\ }\href {\doibase
  10.1103/PhysRevA.101.032104} {\bibfield  {journal} {\bibinfo  {journal}
  {Phys. Rev. A.}\ }\textbf {\bibinfo {volume} {101}},\ \bibinfo {pages}
  {032104} (\bibinfo {year} {2020})}\BibitemShut {NoStop}%
\bibitem [{\citenamefont {Fl{\"a}schner}\ \emph {et~al.}(2018)\citenamefont
  {Fl{\"a}schner}, \citenamefont {Vogel}, \citenamefont {Tarnowski},
  \citenamefont {Rem}, \citenamefont {L{\"u}hmann}, \citenamefont {Heyl},
  \citenamefont {Budich}, \citenamefont {Mathey}, \citenamefont {Sengstock},\
  and\ \citenamefont {Weitenberg}}]{Flaschner:2018aa}%
  \BibitemOpen
  \bibfield  {author} {\bibinfo {author} {\bibfnamefont {N.}~\bibnamefont
  {Fl{\"a}schner}}, \bibinfo {author} {\bibfnamefont {D.}~\bibnamefont
  {Vogel}}, \bibinfo {author} {\bibfnamefont {M.}~\bibnamefont {Tarnowski}},
  \bibinfo {author} {\bibfnamefont {B.~S.}\ \bibnamefont {Rem}}, \bibinfo
  {author} {\bibfnamefont {D.~S.}\ \bibnamefont {L{\"u}hmann}}, \bibinfo
  {author} {\bibfnamefont {M.}~\bibnamefont {Heyl}}, \bibinfo {author}
  {\bibfnamefont {J.~C.}\ \bibnamefont {Budich}}, \bibinfo {author}
  {\bibfnamefont {L.}~\bibnamefont {Mathey}}, \bibinfo {author} {\bibfnamefont
  {K.}~\bibnamefont {Sengstock}}, \ and\ \bibinfo {author} {\bibfnamefont
  {C.}~\bibnamefont {Weitenberg}},\ }\href {\doibase 10.1038/s41567-017-0013-8}
  {\bibfield  {journal} {\bibinfo  {journal} {Nat. Phys.}\ }\textbf {\bibinfo
  {volume} {14}},\ \bibinfo {pages} {265} (\bibinfo {year} {2018})}\BibitemShut
  {NoStop}%
\bibitem [{\citenamefont {Hu}\ and\ \citenamefont
  {Zhao}(2020)}]{HuHaiping2020}%
  \BibitemOpen
  \bibfield  {author} {\bibinfo {author} {\bibfnamefont {H.}~\bibnamefont
  {Hu}}\ and\ \bibinfo {author} {\bibfnamefont {E.}~\bibnamefont {Zhao}},\
  }\href {\doibase 10.1103/PhysRevLett.124.160402} {\bibfield  {journal}
  {\bibinfo  {journal} {Phys. Rev. Lett.}\ }\textbf {\bibinfo {volume} {124}},\
  \bibinfo {pages} {160402} (\bibinfo {year} {2020})}\BibitemShut {NoStop}%
\bibitem [{\citenamefont {Hu}\ \emph {et~al.}(2020{\natexlab{b}})\citenamefont
  {Hu}, \citenamefont {Yang},\ and\ \citenamefont
  {Zhao}}]{PhysRevB.101.155131}%
  \BibitemOpen
  \bibfield  {author} {\bibinfo {author} {\bibfnamefont {H.}~\bibnamefont
  {Hu}}, \bibinfo {author} {\bibfnamefont {C.}~\bibnamefont {Yang}}, \ and\
  \bibinfo {author} {\bibfnamefont {E.}~\bibnamefont {Zhao}},\ }\href {\doibase
  10.1103/PhysRevB.101.155131} {\bibfield  {journal} {\bibinfo  {journal}
  {Phys. Rev. B}\ }\textbf {\bibinfo {volume} {101}},\ \bibinfo {pages}
  {155131} (\bibinfo {year} {2020}{\natexlab{b}})}\BibitemShut {NoStop}%
\bibitem [{\citenamefont {Barnett}\ \emph {et~al.}(2012)\citenamefont
  {Barnett}, \citenamefont {Boyd},\ and\ \citenamefont
  {Galitski}}]{PhysRevLett.109.235308}%
  \BibitemOpen
  \bibfield  {author} {\bibinfo {author} {\bibfnamefont {R.}~\bibnamefont
  {Barnett}}, \bibinfo {author} {\bibfnamefont {G.~R.}\ \bibnamefont {Boyd}}, \
  and\ \bibinfo {author} {\bibfnamefont {V.}~\bibnamefont {Galitski}},\ }\href
  {\doibase 10.1103/PhysRevLett.109.235308} {\bibfield  {journal} {\bibinfo
  {journal} {Phys. Rev. Lett.}\ }\textbf {\bibinfo {volume} {109}},\ \bibinfo
  {pages} {235308} (\bibinfo {year} {2012})}\BibitemShut {NoStop}%
\bibitem [{\citenamefont {Jak{\'o}bczyk}\ and\ \citenamefont
  {Siennicki}(2001)}]{JAKOBCZYK2001383}%
  \BibitemOpen
  \bibfield  {author} {\bibinfo {author} {\bibfnamefont {L.}~\bibnamefont
  {Jak{\'o}bczyk}}\ and\ \bibinfo {author} {\bibfnamefont {M.}~\bibnamefont
  {Siennicki}},\ }\href {\doibase
  https://doi.org/10.1016/S0375-9601(01)00455-8} {\bibfield  {journal}
  {\bibinfo  {journal} {Physics Letters A}\ }\textbf {\bibinfo {volume}
  {286}},\ \bibinfo {pages} {383} (\bibinfo {year} {2001})}\BibitemShut
  {NoStop}%
\bibitem [{\citenamefont {Zyczkowski}\ and\ \citenamefont
  {Sommers}(2003)}]{Zyczkowski2003}%
  \BibitemOpen
  \bibfield  {author} {\bibinfo {author} {\bibfnamefont {K.}~\bibnamefont
  {Zyczkowski}}\ and\ \bibinfo {author} {\bibfnamefont {H.-J.}\ \bibnamefont
  {Sommers}},\ }\href {\doibase 10.1088/0305-4470/36/39/310} {\bibfield
  {journal} {\bibinfo  {journal} {Journal of Physics A: Mathematical and
  General}\ }\textbf {\bibinfo {volume} {36}},\ \bibinfo {pages} {10115}
  (\bibinfo {year} {2003})}\BibitemShut {NoStop}%
\bibitem [{\citenamefont {Kimura}(2003)}]{KIMURA2003339}%
  \BibitemOpen
  \bibfield  {author} {\bibinfo {author} {\bibfnamefont {G.}~\bibnamefont
  {Kimura}},\ }\href {\doibase https://doi.org/10.1016/S0375-9601(03)00941-1}
  {\bibfield  {journal} {\bibinfo  {journal} {Physics Letters A}\ }\textbf
  {\bibinfo {volume} {314}},\ \bibinfo {pages} {339} (\bibinfo {year}
  {2003})}\BibitemShut {NoStop}%
\bibitem [{\citenamefont {Byrd}\ and\ \citenamefont
  {Khaneja}(2003)}]{PhysRevA.68.062322}%
  \BibitemOpen
  \bibfield  {author} {\bibinfo {author} {\bibfnamefont {M.~S.}\ \bibnamefont
  {Byrd}}\ and\ \bibinfo {author} {\bibfnamefont {N.}~\bibnamefont {Khaneja}},\
  }\href {\doibase 10.1103/PhysRevA.68.062322} {\bibfield  {journal} {\bibinfo
  {journal} {Phys. Rev. A}\ }\textbf {\bibinfo {volume} {68}},\ \bibinfo
  {pages} {062322} (\bibinfo {year} {2003})}\BibitemShut {NoStop}%
\bibitem [{\citenamefont {Graf}\ and\ \citenamefont
  {Pi\'echon}(2021)}]{PhysRevB.104.085114}%
  \BibitemOpen
  \bibfield  {author} {\bibinfo {author} {\bibfnamefont {A.}~\bibnamefont
  {Graf}}\ and\ \bibinfo {author} {\bibfnamefont {F.}~\bibnamefont
  {Pi\'echon}},\ }\href {\doibase 10.1103/PhysRevB.104.085114} {\bibfield
  {journal} {\bibinfo  {journal} {Phys. Rev. B}\ }\textbf {\bibinfo {volume}
  {104}},\ \bibinfo {pages} {085114} (\bibinfo {year} {2021})}\BibitemShut
  {NoStop}%
\bibitem [{\citenamefont {Kemp}\ \emph {et~al.}(2022)\citenamefont {Kemp},
  \citenamefont {Cooper},\ and\ \citenamefont
  {\"Unal}}]{PhysRevResearch.4.023120}%
  \BibitemOpen
  \bibfield  {author} {\bibinfo {author} {\bibfnamefont {C.~J.~D.}\
  \bibnamefont {Kemp}}, \bibinfo {author} {\bibfnamefont {N.~R.}\ \bibnamefont
  {Cooper}}, \ and\ \bibinfo {author} {\bibfnamefont {F.~N.}\ \bibnamefont
  {\"Unal}},\ }\href {\doibase 10.1103/PhysRevResearch.4.023120} {\bibfield
  {journal} {\bibinfo  {journal} {Phys. Rev. Res.}\ }\textbf {\bibinfo {volume}
  {4}},\ \bibinfo {pages} {023120} (\bibinfo {year} {2022})}\BibitemShut
  {NoStop}%
\bibitem [{\citenamefont {Hatsugai}(1993)}]{PhysRevLett.71.3697}%
  \BibitemOpen
  \bibfield  {author} {\bibinfo {author} {\bibfnamefont {Y.}~\bibnamefont
  {Hatsugai}},\ }\href {\doibase 10.1103/PhysRevLett.71.3697} {\bibfield
  {journal} {\bibinfo  {journal} {Phys. Rev. Lett.}\ }\textbf {\bibinfo
  {volume} {71}},\ \bibinfo {pages} {3697} (\bibinfo {year}
  {1993})}\BibitemShut {NoStop}%
\bibitem [{\citenamefont {Fukui}(2020)}]{PhysRevResearch.2.043136}%
  \BibitemOpen
  \bibfield  {author} {\bibinfo {author} {\bibfnamefont {T.}~\bibnamefont
  {Fukui}},\ }\href {\doibase 10.1103/PhysRevResearch.2.043136} {\bibfield
  {journal} {\bibinfo  {journal} {Phys. Rev. Res.}\ }\textbf {\bibinfo {volume}
  {2}},\ \bibinfo {pages} {043136} (\bibinfo {year} {2020})}\BibitemShut
  {NoStop}%
\bibitem [{\citenamefont {Hashimoto}\ \emph {et~al.}(2017)\citenamefont
  {Hashimoto}, \citenamefont {Wu},\ and\ \citenamefont
  {Kimura}}]{PhysRevB.95.165443}%
  \BibitemOpen
  \bibfield  {author} {\bibinfo {author} {\bibfnamefont {K.}~\bibnamefont
  {Hashimoto}}, \bibinfo {author} {\bibfnamefont {X.}~\bibnamefont {Wu}}, \
  and\ \bibinfo {author} {\bibfnamefont {T.}~\bibnamefont {Kimura}},\ }\href
  {\doibase 10.1103/PhysRevB.95.165443} {\bibfield  {journal} {\bibinfo
  {journal} {Phys. Rev. B}\ }\textbf {\bibinfo {volume} {95}},\ \bibinfo
  {pages} {165443} (\bibinfo {year} {2017})}\BibitemShut {NoStop}%
\bibitem [{\citenamefont {Davoyan}\ \emph {et~al.}(2024)\citenamefont
  {Davoyan}, \citenamefont {Jankowski}, \citenamefont {Bouhon},\ and\
  \citenamefont {Slager}}]{PhysRevB.109.165125}%
  \BibitemOpen
  \bibfield  {author} {\bibinfo {author} {\bibfnamefont {Z.}~\bibnamefont
  {Davoyan}}, \bibinfo {author} {\bibfnamefont {W.~J.}\ \bibnamefont
  {Jankowski}}, \bibinfo {author} {\bibfnamefont {A.}~\bibnamefont {Bouhon}}, \
  and\ \bibinfo {author} {\bibfnamefont {R.-J.}\ \bibnamefont {Slager}},\
  }\href {\doibase 10.1103/PhysRevB.109.165125} {\bibfield  {journal} {\bibinfo
   {journal} {Phys. Rev. B}\ }\textbf {\bibinfo {volume} {109}},\ \bibinfo
  {pages} {165125} (\bibinfo {year} {2024})}\BibitemShut {NoStop}%
\bibitem [{\citenamefont {Ma\~nes}(2012)}]{PhysRevB.85.155118}%
  \BibitemOpen
  \bibfield  {author} {\bibinfo {author} {\bibfnamefont {J.~L.}\ \bibnamefont
  {Ma\~nes}},\ }\href {\doibase 10.1103/PhysRevB.85.155118} {\bibfield
  {journal} {\bibinfo  {journal} {Phys. Rev. B}\ }\textbf {\bibinfo {volume}
  {85}},\ \bibinfo {pages} {155118} (\bibinfo {year} {2012})}\BibitemShut
  {NoStop}%
\bibitem [{\citenamefont {Liang}\ and\ \citenamefont
  {Yu}(2016)}]{PhysRevB.93.045113}%
  \BibitemOpen
  \bibfield  {author} {\bibinfo {author} {\bibfnamefont {L.}~\bibnamefont
  {Liang}}\ and\ \bibinfo {author} {\bibfnamefont {Y.}~\bibnamefont {Yu}},\
  }\href {\doibase 10.1103/PhysRevB.93.045113} {\bibfield  {journal} {\bibinfo
  {journal} {Phys. Rev. B}\ }\textbf {\bibinfo {volume} {93}},\ \bibinfo
  {pages} {045113} (\bibinfo {year} {2016})}\BibitemShut {NoStop}%
\bibitem [{\citenamefont {Bradlyn}\ \emph {et~al.}(2016)\citenamefont
  {Bradlyn}, \citenamefont {Cano}, \citenamefont {Wang}, \citenamefont
  {Vergniory}, \citenamefont {Felser}, \citenamefont {Cava},\ and\
  \citenamefont {Bernevig}}]{BradlynScience2016}%
  \BibitemOpen
  \bibfield  {author} {\bibinfo {author} {\bibfnamefont {B.}~\bibnamefont
  {Bradlyn}}, \bibinfo {author} {\bibfnamefont {J.}~\bibnamefont {Cano}},
  \bibinfo {author} {\bibfnamefont {Z.}~\bibnamefont {Wang}}, \bibinfo {author}
  {\bibfnamefont {M.~G.}\ \bibnamefont {Vergniory}}, \bibinfo {author}
  {\bibfnamefont {C.}~\bibnamefont {Felser}}, \bibinfo {author} {\bibfnamefont
  {R.~J.}\ \bibnamefont {Cava}}, \ and\ \bibinfo {author} {\bibfnamefont
  {B.~A.}\ \bibnamefont {Bernevig}},\ }\href {\doibase 10.1126/science.aaf5037}
  {\bibfield  {journal} {\bibinfo  {journal} {Science}\ }\textbf {\bibinfo
  {volume} {353}},\ \bibinfo {pages} {aaf5037} (\bibinfo {year} {2016})},\
  \Eprint
  {http://arxiv.org/abs/https://www.science.org/doi/pdf/10.1126/science.aaf5037}
  {https://www.science.org/doi/pdf/10.1126/science.aaf5037} \BibitemShut
  {NoStop}%
\bibitem [{\citenamefont {Tang}\ \emph {et~al.}(2017)\citenamefont {Tang},
  \citenamefont {Zhou},\ and\ \citenamefont {Zhang}}]{PhysRevLett.119.206402}%
  \BibitemOpen
  \bibfield  {author} {\bibinfo {author} {\bibfnamefont {P.}~\bibnamefont
  {Tang}}, \bibinfo {author} {\bibfnamefont {Q.}~\bibnamefont {Zhou}}, \ and\
  \bibinfo {author} {\bibfnamefont {S.-C.}\ \bibnamefont {Zhang}},\ }\href
  {\doibase 10.1103/PhysRevLett.119.206402} {\bibfield  {journal} {\bibinfo
  {journal} {Phys. Rev. Lett.}\ }\textbf {\bibinfo {volume} {119}},\ \bibinfo
  {pages} {206402} (\bibinfo {year} {2017})}\BibitemShut {NoStop}%
\bibitem [{\citenamefont {Chang}\ \emph {et~al.}(2017)\citenamefont {Chang},
  \citenamefont {Xu}, \citenamefont {Wieder}, \citenamefont {Sanchez},
  \citenamefont {Huang}, \citenamefont {Belopolski}, \citenamefont {Chang},
  \citenamefont {Zhang}, \citenamefont {Bansil}, \citenamefont {Lin},\ and\
  \citenamefont {Hasan}}]{PhysRevLett.119.206401}%
  \BibitemOpen
  \bibfield  {author} {\bibinfo {author} {\bibfnamefont {G.}~\bibnamefont
  {Chang}}, \bibinfo {author} {\bibfnamefont {S.-Y.}\ \bibnamefont {Xu}},
  \bibinfo {author} {\bibfnamefont {B.~J.}\ \bibnamefont {Wieder}}, \bibinfo
  {author} {\bibfnamefont {D.~S.}\ \bibnamefont {Sanchez}}, \bibinfo {author}
  {\bibfnamefont {S.-M.}\ \bibnamefont {Huang}}, \bibinfo {author}
  {\bibfnamefont {I.}~\bibnamefont {Belopolski}}, \bibinfo {author}
  {\bibfnamefont {T.-R.}\ \bibnamefont {Chang}}, \bibinfo {author}
  {\bibfnamefont {S.}~\bibnamefont {Zhang}}, \bibinfo {author} {\bibfnamefont
  {A.}~\bibnamefont {Bansil}}, \bibinfo {author} {\bibfnamefont
  {H.}~\bibnamefont {Lin}}, \ and\ \bibinfo {author} {\bibfnamefont {M.~Z.}\
  \bibnamefont {Hasan}},\ }\href {\doibase 10.1103/PhysRevLett.119.206401}
  {\bibfield  {journal} {\bibinfo  {journal} {Phys. Rev. Lett.}\ }\textbf
  {\bibinfo {volume} {119}},\ \bibinfo {pages} {206401} (\bibinfo {year}
  {2017})}\BibitemShut {NoStop}%
\bibitem [{\citenamefont {Schr{\"o}ter}\ \emph {et~al.}(2019)\citenamefont
  {Schr{\"o}ter}, \citenamefont {Pei}, \citenamefont {Vergniory}, \citenamefont
  {Sun}, \citenamefont {Manna}, \citenamefont {de~Juan}, \citenamefont
  {Krieger}, \citenamefont {S{\"u}ss}, \citenamefont {Schmidt}, \citenamefont
  {Dudin}, \citenamefont {Bradlyn}, \citenamefont {Kim}, \citenamefont
  {Schmitt}, \citenamefont {Cacho}, \citenamefont {Felser}, \citenamefont
  {Strocov},\ and\ \citenamefont {Chen}}]{Schroter:2019aa}%
  \BibitemOpen
  \bibfield  {author} {\bibinfo {author} {\bibfnamefont {N.~B.~M.}\
  \bibnamefont {Schr{\"o}ter}}, \bibinfo {author} {\bibfnamefont
  {D.}~\bibnamefont {Pei}}, \bibinfo {author} {\bibfnamefont {M.~G.}\
  \bibnamefont {Vergniory}}, \bibinfo {author} {\bibfnamefont {Y.}~\bibnamefont
  {Sun}}, \bibinfo {author} {\bibfnamefont {K.}~\bibnamefont {Manna}}, \bibinfo
  {author} {\bibfnamefont {F.}~\bibnamefont {de~Juan}}, \bibinfo {author}
  {\bibfnamefont {J.~A.}\ \bibnamefont {Krieger}}, \bibinfo {author}
  {\bibfnamefont {V.}~\bibnamefont {S{\"u}ss}}, \bibinfo {author}
  {\bibfnamefont {M.}~\bibnamefont {Schmidt}}, \bibinfo {author} {\bibfnamefont
  {P.}~\bibnamefont {Dudin}}, \bibinfo {author} {\bibfnamefont
  {B.}~\bibnamefont {Bradlyn}}, \bibinfo {author} {\bibfnamefont {T.~K.}\
  \bibnamefont {Kim}}, \bibinfo {author} {\bibfnamefont {T.}~\bibnamefont
  {Schmitt}}, \bibinfo {author} {\bibfnamefont {C.}~\bibnamefont {Cacho}},
  \bibinfo {author} {\bibfnamefont {C.}~\bibnamefont {Felser}}, \bibinfo
  {author} {\bibfnamefont {V.~N.}\ \bibnamefont {Strocov}}, \ and\ \bibinfo
  {author} {\bibfnamefont {Y.}~\bibnamefont {Chen}},\ }\href {\doibase
  10.1038/s41567-019-0511-y} {\bibfield  {journal} {\bibinfo  {journal} {Nature
  Physics}\ }\textbf {\bibinfo {volume} {15}},\ \bibinfo {pages} {759}
  (\bibinfo {year} {2019})}\BibitemShut {NoStop}%
\bibitem [{\citenamefont {Robredo}\ \emph {et~al.}(2024)\citenamefont
  {Robredo}, \citenamefont {Schr{\"o}eter}, \citenamefont {Felser},
  \citenamefont {Cano}, \citenamefont {Bradlyn},\ and\ \citenamefont
  {Vergniory}}]{Robredo:2024aa}%
  \BibitemOpen
  \bibfield  {author} {\bibinfo {author} {\bibfnamefont {I.}~\bibnamefont
  {Robredo}}, \bibinfo {author} {\bibfnamefont {N.}~\bibnamefont
  {Schr{\"o}eter}}, \bibinfo {author} {\bibfnamefont {C.}~\bibnamefont
  {Felser}}, \bibinfo {author} {\bibfnamefont {J.}~\bibnamefont {Cano}},
  \bibinfo {author} {\bibfnamefont {B.}~\bibnamefont {Bradlyn}}, \ and\
  \bibinfo {author} {\bibfnamefont {M.~G.}\ \bibnamefont {Vergniory}},\ }\href
  {https://arxiv.org/pdf/2404.17539v1.pdf} {\  (\bibinfo {year} {2024})},\
  \Eprint {http://arxiv.org/abs/2404.17539} {2404.17539} \BibitemShut {NoStop}%
\bibitem [{\citenamefont {Titum}\ \emph {et~al.}(2016)\citenamefont {Titum},
  \citenamefont {Berg}, \citenamefont {Rudner}, \citenamefont {Refael},\ and\
  \citenamefont {Lindner}}]{PhysRevX.6.021013}%
  \BibitemOpen
  \bibfield  {author} {\bibinfo {author} {\bibfnamefont {P.}~\bibnamefont
  {Titum}}, \bibinfo {author} {\bibfnamefont {E.}~\bibnamefont {Berg}},
  \bibinfo {author} {\bibfnamefont {M.~S.}\ \bibnamefont {Rudner}}, \bibinfo
  {author} {\bibfnamefont {G.}~\bibnamefont {Refael}}, \ and\ \bibinfo {author}
  {\bibfnamefont {N.~H.}\ \bibnamefont {Lindner}},\ }\href {\doibase
  10.1103/PhysRevX.6.021013} {\bibfield  {journal} {\bibinfo  {journal} {Phys.
  Rev. X}\ }\textbf {\bibinfo {volume} {6}},\ \bibinfo {pages} {021013}
  (\bibinfo {year} {2016})}\BibitemShut {NoStop}%
\bibitem [{\citenamefont {Bessho}\ and\ \citenamefont
  {Sato}(2021)}]{PhysRevLett.127.196404}%
  \BibitemOpen
  \bibfield  {author} {\bibinfo {author} {\bibfnamefont {T.}~\bibnamefont
  {Bessho}}\ and\ \bibinfo {author} {\bibfnamefont {M.}~\bibnamefont {Sato}},\
  }\href {\doibase 10.1103/PhysRevLett.127.196404} {\bibfield  {journal}
  {\bibinfo  {journal} {Phys. Rev. Lett.}\ }\textbf {\bibinfo {volume} {127}},\
  \bibinfo {pages} {196404} (\bibinfo {year} {2021})}\BibitemShut {NoStop}%
\bibitem [{\citenamefont {Nielsen}\ and\ \citenamefont
  {Ninomiya}(1981{\natexlab{a}})}]{NIELSEN198120}%
  \BibitemOpen
  \bibfield  {author} {\bibinfo {author} {\bibfnamefont {H.}~\bibnamefont
  {Nielsen}}\ and\ \bibinfo {author} {\bibfnamefont {M.}~\bibnamefont
  {Ninomiya}},\ }\href {\doibase https://doi.org/10.1016/0550-3213(81)90361-8}
  {\bibfield  {journal} {\bibinfo  {journal} {Nuclear Physics B}\ }\textbf
  {\bibinfo {volume} {185}},\ \bibinfo {pages} {20} (\bibinfo {year}
  {1981}{\natexlab{a}})}\BibitemShut {NoStop}%
\bibitem [{\citenamefont {Nielsen}\ and\ \citenamefont
  {Ninomiya}(1981{\natexlab{b}})}]{NIELSEN1981173}%
  \BibitemOpen
  \bibfield  {author} {\bibinfo {author} {\bibfnamefont {H.}~\bibnamefont
  {Nielsen}}\ and\ \bibinfo {author} {\bibfnamefont {M.}~\bibnamefont
  {Ninomiya}},\ }\href {\doibase https://doi.org/10.1016/0550-3213(81)90524-1}
  {\bibfield  {journal} {\bibinfo  {journal} {Nuclear Physics B}\ }\textbf
  {\bibinfo {volume} {193}},\ \bibinfo {pages} {173} (\bibinfo {year}
  {1981}{\natexlab{b}})}\BibitemShut {NoStop}%
\bibitem [{\citenamefont {\"Unal}\ \emph {et~al.}(2019)\citenamefont {\"Unal},
  \citenamefont {Seradjeh},\ and\ \citenamefont
  {Eckardt}}]{PhysRevLett.122.253601}%
  \BibitemOpen
  \bibfield  {author} {\bibinfo {author} {\bibfnamefont {F.~N.}\ \bibnamefont
  {\"Unal}}, \bibinfo {author} {\bibfnamefont {B.}~\bibnamefont {Seradjeh}}, \
  and\ \bibinfo {author} {\bibfnamefont {A.}~\bibnamefont {Eckardt}},\ }\href
  {\doibase 10.1103/PhysRevLett.122.253601} {\bibfield  {journal} {\bibinfo
  {journal} {Phys. Rev. Lett.}\ }\textbf {\bibinfo {volume} {122}},\ \bibinfo
  {pages} {253601} (\bibinfo {year} {2019})}\BibitemShut {NoStop}%
\bibitem [{\citenamefont {Si}(2005)}]{Si2005}%
  \BibitemOpen
  \bibfield  {author} {\bibinfo {author} {\bibfnamefont {T.}~\bibnamefont
  {Si}},\ }\href {\doibase 10.1063/1.2137721} {\bibfield  {journal} {\bibinfo
  {journal} {Journal of Mathematical Physics}\ }\textbf {\bibinfo {volume}
  {46}},\ \bibinfo {pages} {122301} (\bibinfo {year} {2005})},\ \Eprint
  {http://arxiv.org/abs/https://pubs.aip.org/aip/jmp/article-pdf/doi/10.1063/1.2137721/15867792/122301\_1\_online.pdf}
  {https://pubs.aip.org/aip/jmp/article-pdf/doi/10.1063/1.2137721/15867792/122301\_1\_online.pdf}
  \BibitemShut {NoStop}%
\end{thebibliography}%
\end{document}